\documentclass[final,12pt,3p]{elsarticle}

\usepackage{graphicx}
\usepackage{amssymb}
\usepackage{amsthm}
\usepackage{amsmath}
\usepackage{algorithm}
\usepackage{algorithmic}
\usepackage{tikz}
\usepackage{lineno}
\usepackage{verbatim} 
\usepackage{hyperref}
 \usepackage{xcolor}
\usepackage{caption}
\usetikzlibrary{shapes}
\usepackage{subcaption}
\usepackage{geometry}
\usepackage{mathrsfs} 
\usepackage{array}
\usepackage{multirow}
\usepackage{booktabs} 
\usepackage{adjustbox}
\usepackage[title]{appendix}

\biboptions{sort&compress}
\hypersetup{
colorlinks=true,
linkcolor=red,
filecolor=green,
urlcolor=green,
anchorcolor=green,
citecolor=cyan,
pdftitle={Overleaf Example},
pdfpagemode=FullScreen,
}
\newtheorem{remark}{Remark}
\newtheorem{corollary}{Corollary}

\usepackage[figcolor=white]{todonotes}

\begin{document}

\begin{frontmatter}

\begin{tikzpicture}[remember picture,overlay]
	\node[anchor=north east,inner sep=20pt] at (current page.north east)
	{\includegraphics[scale=0.2]{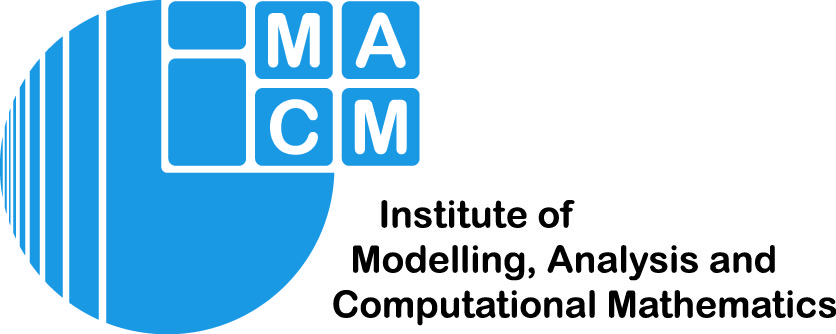}};
\end{tikzpicture}

 \title{A Koopman-PINN Framework for Epidemic Models: \\ Parameter Inference and Forecasting}
%

\author[BUW,MAIS]{Achraf Zinihi}
\ead{a.zinihi@edu.umi.ac.ma} 

\author[BUW]{Matthias Ehrhardt\corref{Corr}}
\cortext[Corr]{Corresponding author}
\ead{ehrhardt@uni-wuppertal.de}

\author[MAIS]{Moulay Rchid Sidi Ammi}
\ead{rachidsidiammi@yahoo.fr}

\address[BUW]{University of Wuppertal, Applied and Computational Mathematics,\\
Gaußstrasse 20, 42119 Wuppertal, Germany}

\address[MAIS]{Department of Mathematics, AMNEA Group, Faculty of Sciences and Techniques,\\
Moulay Ismail University of Meknes, Errachidia 52000, Morocco}


\begin{abstract}
We propose a Koopman-enhanced physics-informed neural network (K--PINN) framework for parameter inference and forecasting in nonlinear epidemic models. This method combines Koopman operator theory and physics-informed learning. It maps epidemic states into a latent observable space where the dynamics evolve approximately linearly while satisfying the governing epidemic equations through automatic differentiation. This integration improves interpretability, parameter identifiability, and long-term predictive stability.

We apply the proposed framework to a normalized SEIRSD epidemic model and evaluate it using synthetic monkeypox (Mpox) data and real-world datasets from Germany, Morocco, and Sweden for the SARS-CoV-2 virus. Synthetic trajectories are generated using a structure-preserving, nonstandard finite difference scheme to ensure reliable training data. Numerical results demonstrate that K--PINN achieves more accurate parameter estimation, trajectory reconstruction, and long-term forecasting than classical PINNs and Koopman-EDMD approaches.

These results suggest that K--PINN is an effective machine learning framework for epidemic modeling that can be extended to more complex systems.

\end{abstract}






\end{frontmatter}


\section{Introduction}\label{S1}
Mathematical epidemiology is fundamental to understanding, predicting, and controlling the spread of infectious diseases. 
Classical compartmental models have been widely used to describe the transmission dynamics of diseases such as influenza, Ebola, and SARS-CoV-2. 
Due to their interpretability and mathematical structure, these models offer valuable insights into epidemic propagation, stability analysis, intervention strategies, and long-term forecasting. 
However, practical epidemiological modeling is challenging due to incomplete observations, noisy and sparse data, underreporting, and the nonlinear nature of epidemic dynamics. 
These difficulties often complicate parameter estimation, hidden-state reconstruction, and reliable forecasting, especially when real epidemiological data is limited or uncertain.

In recent years, physics-informed neural networks (PINNs) have emerged as a powerful machine learning framework in physics for solving forward and inverse problems governed by differential equations. 
PINNs were introduced by Raissi, Perdikaris, and Karniadakis in 2019 \cite{Raissi2019PINN} and integrate physical laws directly into neural network training through automatic differentiation. This allows the learned solutions to satisfy the governing differential equations while simultaneously fitting observational data. 
This approach has demonstrated strong capabilities in parameter estimation, hidden-state identification, and forecasting for nonlinear dynamical systems. 
In epidemiological applications, PINNs have shown promising performance in inferring epidemic parameters and in data-driven reconstruction from sparse observations. 
However, classical PINN frameworks may suffer from training instability, parameter identifiability issues, and reduced long-term predictive accuracy when epidemic dynamics become highly nonlinear or observational data is scarce.

In parallel with these developments, Koopman operator theory has attracted increasing attention in the fields of dynamical systems and data-driven modeling. 
Originally introduced by Koopman \cite{Koopman1931} and further developed by Mezić \cite{Mezi2005}, 
the Koopman framework provides a linear operator-theoretic representation of nonlinear dynamical systems by acting on observable functions instead of state variables directly. 
This enables nonlinear systems to be analyzed using linear spectral techniques and operator theory. 
Data-driven Koopman approaches, including \textit{dynamic mode decomposition} (DMD) and \textit{extended dynamic mode decomposition} (EDMD), 
have been widely applied to prediction, model reduction, and system identification  \cite{Korda2018, Proctor2016}.  
Recently, deep learning architectures have been introduced to automatically learn Koopman-invariant latent representations and nonlinear observable mappings \cite{Lusch2018, Takeishi2017, Otto2019, Champion2019}. 
However, many existing approaches rely on handcrafted observables or purely data-driven embeddings that may not preserve the physical and epidemiological structure of the underlying system.

Several recent works have explored advanced mathematical and computational methodologies for epidemic modeling, motivated by these challenges.
These works include \cite{Zinihi2025FDE, Zinihi2025Spatiotemporal, Zinihi2026CAPINN, Zinihi2025MC, Zinihi2025PINN, Zinihi2026Koopman}. 
For instance, the authors' studies \cite{Zinihi2025MC, Zinihi2026CAPINN} have developed advanced computational frameworks for analyzing complex epidemic dynamics in spatially distributed and interconnected population systems. 
In \cite{Zinihi2026CAPINN}, a constraint-aware PINN approach was proposed for reaction-diffusion epidemic models. 
In another study \cite{Zinihi2025MC}, a hybrid multi-city epidemic modeling strategy was introduced, combining topology-based transportation networks, self-consistent coupling parameters, and distance-dependent temporal migration delays to capture the spread of infectious diseases across interconnected urban centers.
In \cite{Zinihi2025PINN}, a PINN framework was developed for fractional-order epidemic models where the Caputo fractional derivative was incorporated into the learning process to estimate epidemiological parameters and the fractional memory order simultaneously. 
In another study conducted by Zinihi, Ehrhardt, and Sidi Ammi \cite{Zinihi2026Koopman}, 
an epidemic model incorporating waning immunity and disease-induced mortality was developed and analyzed within the Koopman operator framework. 
The study demonstrated the ability of Koopman-based approaches to capture nonlinear epidemic dynamics and accurately predict important outbreak features, such as peak infection behavior, using 
EDMD together with epidemiologically informed observables.

Based on these developments, this paper proposes a \textit{Koopman-enhanced physics-informed neural network} (K--PINN) framework for nonlinear epidemic systems. 
To the best of our knowledge, this is the first attempt to combine Koopman-based latent linearization and physics-informed learning for epidemic compartmental models. 
This methodology leverages the advantages of Koopman analysis, such as operator theory, and the capabilities of PINNs, such as equation enforcement, through a neural lifting architecture. 
This architecture maps epidemic states into a Koopman-compatible latent space with approximately linear evolution. 
The governing epidemic equations are simultaneously enforced through a physics-informed loss function based on automatic differentiation. 
This framework improves parameter identifiability, latent stability, and long-term forecasting while preserving the mechanistic structure of epidemic dynamics.

The main contributions of this work are as follows:
First, we propose a Koopman-enhanced physics-informed neural network (K--PINN) framework for nonlinear epidemic systems. 
This framework uses a neural lifting architecture to automatically learn Koopman observables and map epidemic dynamics into a latent representation with approximately linear evolution. 
Meanwhile, the governing epidemic equations are enforced through a physics-informed optimization procedure based on automatic differentiation. 
This enables a robust inverse-learning strategy for parameter estimation and forecasting under sparse and noisy observations. We further benchmark the proposed methodology against classical Koopman-EDMD (K--EDMD) and standard PINN approaches and validate it on SEIRSD epidemic dynamics using synthetic and real epidemiological datasets.

In addition, structure-preserving numerical schemes, such as \textit{nonstandard finite difference} (NSFD) methods, provide stable and positivity-preserving epidemic simulations. This makes them particularly suitable for generating reliable synthetic training data \cite{Ehrhardt2013, Zinihi2025NSFD}.

The remainder of this paper is organized as follows. 
Section~\ref{S2} introduces the proposed epidemic model together with its formulation, normalization, and well-posedness analysis. 
Section~\ref{S3} presents the proposed K--PINN architecture and the associated physics-informed loss functions. 
Section~\ref{S4} describes the numerical methodology, training procedure, and synthetic data generation strategy. 
Section~\ref{S5} presents numerical experiments and forecasting results using both synthetic and real epidemiological datasets. 
Finally, Section~\ref{S6} concludes the paper and discusses future research directions.

\section{Proposed Model and Mathematical Formulation}\label{S2}
We consider a compartmental epidemic model that divides the population into five epidemiological classes: susceptible $S(t)$, exposed $E(t)$, infected $I(t)$, recovered $R(t)$, and deceased $D(t)$. Let
\begin{equation*}
    N_L(t) = S(t) + E(t) + I(t) + R(t),
\end{equation*}
denote the \textit{total living population} at time $t$. 
The disease transmission dynamics are described by the following nonlinear SEIRSD system
\begin{equation}\label{E2.1}
\left\{\begin{aligned}
\dot{S}(t) &= -\beta \frac{S(t)I(t)}{N_L(t)} + \alpha R(t), \\
\dot{E}(t) &= \beta \frac{S(t)I(t)}{N_L(t)} - \sigma E(t) , \\
\dot{I}(t) &= \sigma E(t) - (\gamma + \mu)I(t), \\
\dot{R}(t) &= \gamma I(t) - \alpha R(t),\\  
\dot{D}(t) &= \mu I(t),
\end{aligned}\right.
\end{equation}
supplied with the initial conditions
\begin{equation}\label{E2.2}
S(0)=S_0>0, \quad E(0)=E_0>0, \quad I(0)=I_0>0, \quad R(0)=R_0\ge0, \quad D(0)=D_0\ge0.
\end{equation}
The parameter $\beta>0$ represents the effective transmission rate, $\sigma>0$ denotes the progression rate from the exposed to the infected class, $\gamma>0$ is the recovery rate, $\mu>0$ corresponds to the disease-induced mortality rate, and $\alpha \ge0$ characterizes the loss of acquired immunity. 
Recovered individuals return to the susceptible compartment when immunity wanes. Permanent immunity is achieved in the special case $\alpha = 0$.

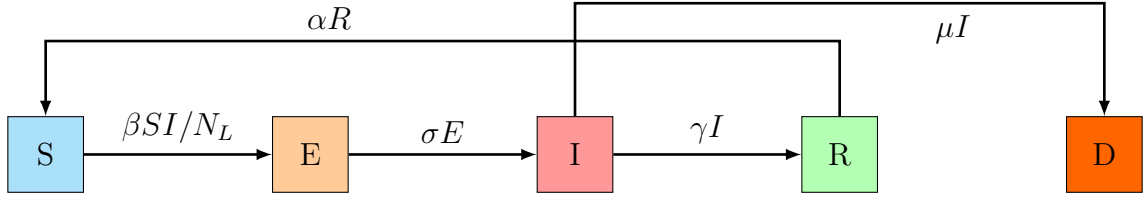
\begin{figure}[H]
\centering
\begin{tikzpicture}[node distance=3.5cm]
\node (S) [rectangle, draw, minimum size=1cm, fill=cyan!30] {S};
\node (E) [rectangle, draw, minimum size=1cm, fill=orange!40, right of=S] {E};
\node (I) [rectangle, draw, minimum size=1cm, fill=red!40, right of=E] {I};
\node (R) [rectangle, draw, minimum size=1cm, fill=green!30, right of=I] {R};
\node (D) [rectangle, draw, minimum size=1cm, fill=orange!80!red, right of=R] {D};
\draw [->, thick, >=latex, line width=1pt] (S) -- (E) node[midway,above]{$\beta S I / N_L$};
\draw [->, thick, >=latex, line width=1pt] (E) -- (I) node[midway,above]{$\sigma E$};
\draw [->, thick, >=latex, line width=1pt] (I) -- (R) node[midway,above]{$\gamma I$};
\draw [<-, thick, >=latex, line width=1pt] (S.north) -- (0,1.5) -- (10.5,1.5) node[midway,above,xshift=-1.5cm]{$\alpha R$} -- (R.north);
\draw [->, thick, >=latex, line width=1pt] (I.north) -- (7,2) -- (14,2) node[midway,below,xshift=1.5cm]{$\mu I$} -- (D.north);
\end{tikzpicture}
\captionof{figure}{Flow diagram of the proposed SEIRSD model.}\label{F1}
\end{figure}

The infection process is governed by the incidence term $\beta S(t)I(t)/N_L(t)$, which assumes frequency-dependent transmission. 
Susceptible individuals become exposed through effective contact with infected individuals. Exposed individuals progress to the infected stage at a rate $\sigma$, and infected individuals either recover or die from the disease at rates $\gamma$ and $\mu$, respectively. 
Recovered individuals may lose immunity and re-enter the susceptible class at rate $\alpha$, while the deceased compartment accumulates disease-induced fatalities and does not participate in disease transmission. 
Figure~\ref{F1} presents a schematic representation of these transitions.
The nonlinear nature of the model~\eqref{E2.1} originates from the bilinear transmission term and the dependence of the incidence rate on the time-varying living population $N_L(t)$. 
Consequently, the model exhibits richer dynamic behavior than classical epidemic models and provides a realistic framework for studying outbreaks in the presence of disease-induced mortality and temporary immunity.

\begin{remark}
The model~\eqref{E2.1} does not incorporate vital dynamics, such as natural births and natural deaths. This simplification is justified because epidemic outbreaks generally occur over time horizons that are significantly shorter than those associated with demographic evolution. 
Under this assumption, population changes are driven solely by disease-related transitions, while the total population remains constant throughout the study period. 
This formulation allows the analysis to focus solely on the mechanisms governing disease transmission, progression, recovery, and mortality.
\end{remark}

To establish the basic dynamical properties of~\eqref{E2.1}, we first consider the \textit{total population} $N(t) = S(t) + E(t) + I(t) + R(t) + D(t)$.
Summing all the equations of~\eqref{E2.1} yields
\begin{equation}\label{eq:conservation}
    \dot{N}(t) = \dot{S}(t) + \dot{E}(t) + \dot{I}(t) + \dot{R}(t) + \dot{D}(t) \stackrel{\eqref{E2.1}}{=} 0.
\end{equation}
Therefore, the total population remains constant throughout the epidemic evolution, i.e.\ $N(t) = N_0$.
As a consequence, all trajectories of system~\eqref{E2.1} remain bounded in the positively invariant set
\begin{equation*}
   \Omega = \Big\{(S,E,I,R,D)\in\mathbb{R}_+^5 \colon S + E + I + R + D = N_0 \Big\}.
\end{equation*}
Furthermore, $N_L(t) = N(t) - D(t)$, so differentiation yields $\dot{N}_L(t) = -\mu I(t)$.
Since $I(t)\ge0$ and $\mu>0$, it follows that $\dot{N}_L(t)\le0$, showing that the living population decreases over time. 
However, since $N_L(0)>0$ and deaths occur continuously, the quantity $N_L(t)$ remains strictly positive for every finite time interval.
Standard existence and uniqueness results apply because the right-hand side of system~\eqref{E2.1} is locally Lipschitz in the biologically feasible region where $N_L(t)>0$. 
Combining local existence with the boundedness and positivity properties established above guarantees global solvability.
A detailed proof of the existence, uniqueness, positivity, and boundedness of solutions can be obtained by following the methodology developed in \cite[pp.~4--6]{Zinihi2025PINN} and \cite[pp.~3--5]{Zinihi2026Koopman}. 
Therefore, only the main result is stated here.
For completeness, the main results are summarized below.
\begin{corollary}\label{C1}
For any admissible parameter set $\beta,\sigma,\gamma,\mu>0$, and $\alpha\ge0$, together with initial conditions that satisfy \eqref{E2.2}, system~\eqref{E2.1} admits a unique global solution that remains nonnegative and bounded for all $t\ge0$.
\end{corollary}

Since the deceased compartment does not directly affect the dynamics of the remaining epidemiological classes, except through the living population, it is convenient to reformulate the model in terms of population fractions. 
We therefore introduce the following dimensionless variables
\begin{equation*}
     s(t) = \frac{S(t)}{N}, \quad e(t) = \frac{E(t)}{N}, \quad i(t) = \frac{I(t)}{N}, \quad r(t) = \frac{R(t)}{N}, \quad d(t) = \frac{D(t)}{N}.
\end{equation*}
Using the conservation property \eqref{eq:conservation}, the living population can be expressed as
\begin{equation*}
     N_L(t) = S(t) + E(t) + I(t) + R(t) = N - D(t) = N \bigl( 1 - d(t) \bigr),
\end{equation*}
Substituting the normalized variables into~\eqref{E2.1} and dividing by $N$ yields
\begin{equation}\label{E2.3}
\left\{\begin{aligned}
\dot{s}(t) &= -\beta \frac{s(t) i(t)}{1 - d(t)} + \alpha r(t), \\
\dot{e}(t) &= \beta \frac{s(t) i(t)}{1 - d(t)} - \sigma e(t), \\
\dot{i}(t) &= \sigma e(t) - (\gamma + \mu) i(t), \\
\dot{r}(t) &= \gamma i(t) - \alpha r(t),
\end{aligned}
\right.
\end{equation}
with 
\begin{equation}\label{E2.4}
    s(0) = s_0, \quad e(0) = e_0, \quad i(0) = i_0, \quad r(0) = r_0, \quad d(0) = d_0,
\end{equation}
where
\begin{equation}\label{E2.5}
     d(t) = 1 - s(t) - e(t) - i(t) - r(t).
\end{equation}
Since $\dot{s} + \dot{e} + \dot{i} + \dot{r} + \dot{d} = 0$, showing that the simplex
\begin{equation*}
     \Delta = \Big\{(s,e,i,r,d)\in\mathbb{R}_+^5 \colon s + e + i + r + d = 1 \ \text{ and } \ d < 1 \Big\},
\end{equation*}
is forward invariant. Therefore, if the initial conditions belong to $\Delta$, then all state variables remain nonnegative and bounded by 1.

The normalized system~\eqref{E2.3}--\eqref{E2.4} serves as the reference model for the subsequent Koopman analysis and the development of the proposed K--PINN framework. 
This formulation offers several important advantages: 
$(i)$ The state variables $s(t)$, $e(t)$, $i(t)$, $r(t)$, and $d(t)$ represent population proportions and are therefore naturally bounded in the interval $[0,1]$. 
This improves numerical conditioning and facilitates the enforcement of positivity and conservation constraints. 
$(ii)$ The use of dimensionless variables eliminates the dependence on the total population size. This allows epidemiological parameters and dynamical behaviors to be interpreted and compared more consistently across different populations and datasets. 
$(iii)$ From a Koopman operator perspective, normalized variables reduce scaling disparities among observables and provide a more homogeneous representation of the dynamics. 
This improves the construction of Koopman-invariant latent spaces and the learning of linear representations of the underlying nonlinear dynamics. 
$(iv)$ The factor $(1-d(t))^{-1}$ explicitly incorporates the effect of disease-induced mortality through the time-varying living population. 
This preserves the epidemiological realism of the transmission process while maintaining a scale-free formulation. 
For these reasons, all subsequent theoretical developments, Koopman formulations, and K--PINN constructions are based on the normalized system~\eqref{E2.3}--\eqref{E2.4}.

\section{Koopman-Enhanced Physics-Informed Neural Networks}\label{S3}
The normalized model introduced in Section~\ref{S2} provides a biologically meaningful and mathematically consistent description of epidemic dynamics. 
However, it remains difficult to estimate unknown epidemiological parameters and forecast long-term behavior from sparse or noisy observations. 
The dynamics are nonlinear, only partially observed, and sensitive to parameter uncertainty. 
Classical numerical fitting methods may capture observed data locally, yet they fail to preserve the underlying epidemic structure. 
Standard neural networks can approximate trajectories but do not respect differential equations.

PINNs improve this situation by embedding the governing equations into the learning process. 
Nevertheless, for strongly nonlinear epidemic systems, PINNs may suffer from optimization stiffness, identifiability issues, and limited extrapolation ability. 
Koopman operator theory offers a complementary perspective by lifting the nonlinear dynamics into a space of observables, in which the evolution is approximately linear. 
This linear latent evolution can improve interpretability and long-term prediction.

We combine these ideas through a K-PINN framework. 
The proposed method uses a single, epidemiologically-informed observable representation; evolves the lifted variables under a finite-dimensional Koopman operator; reconstructs the original epidemic state via a decoder; and enforces the normalized SEIRSD equations through physics-based regularization. 
This produces a unified inverse-learning model that improves accuracy, leverages domain knowledge, and enables robust epidemic reconstruction, parameter inference, and forecasting.

\subsection{Proposed Observable Dictionary}\label{S3.1}
Let
\begin{equation*}
    \mathbf{x}(t) = \big[s(t), \,e(t), \,i(t), \,r(t), \,d(t)\big]^{\top} \in \mathbb{R}^5,
\end{equation*}
be the normalized epidemic state. 
To represent the nonlinear structure of the SEIRSD system~\eqref{E2.3}--\eqref{E2.5} in a lifted space, we define a single observable dictionary
\begin{equation*}
    \Phi(\mathbf{x}) = \big[\phi_1(\mathbf{x}), \,\phi_2(\mathbf{x}), \,\dots, \,\phi_m(\mathbf{x})\big]^{\top}\in\mathbb{R}^m.
\end{equation*}
A suitable epidemiological dictionary may include the basic compartments together with selected nonlinear interaction terms, such as
\begin{equation*}
    \Phi(\mathbf{x}) = \bigl[s, \,e, \,i, \,r, \,d, \,si, \,se, \,sr, \,ei, \,er, \,ir, \,\frac{s i}{1 - d}, \,s^2, \,e^2, \,i^2, \,r^2, \,d^2\bigr]^{\top}.
\end{equation*}
The goal is to provide a compact lifted representation that retains interpretability while capturing the essential nonlinear couplings of the epidemic process.

In the K--PINN framework, the lifting function $\Phi(\mathbf{x})$ determines the observable space in which the nonlinear epidemic dynamics are approximated by a finite-dimensional Koopman operator. 
For the SEIRSD model~\eqref{E2.3}, we augment the set of observables beyond the bilinear transmission term $si$ by introducing several quadratic and cross-product observables, such as $sr, ei, er, ir, s^2, e^2, i^2, r^2$, and $d^2$. 
While only a subset of these observables admits a direct epidemiological interpretation, such as the infection incidence term $si$, the other terms are included to enrich the functional basis and improve the approximation of the underlying Koopman-invariant subspace. 
These higher-order observables provide additional flexibility in representing nonlinear interactions and are widely used in reduced-order and nonlinear dynamical system modeling. 
Furthermore, quadratic terms, such as $s^2$, $e^2$, and $i^2$ can be interpreted as capturing density-dependent or feedback effects in epidemiological dynamics \cite{Zinihi2026Koopman}.
Unlike multi-dictionary K--EDMD studies, the present framework relies on one coherent observable set. This is sufficient because the physics-informed loss and the latent linear dynamics jointly guide the learning process.

\subsection{Latent Koopman Dynamics}\label{S3.2}
The observable vector is mapped to a latent Koopman representation by
\begin{equation*}
    \mathbf{z}(t)=\Phi_{\theta}(\mathbf{x}(t)),
\end{equation*}
where $\Phi_{\theta}$ denotes either a fixed dictionary map or a trainable neural lifting operator. 
The latent state $\mathbf{z}(t)\in\mathbb{R}^m$ is assumed to evolve according to
\begin{equation*}
    \frac{d\mathbf{z}(t)}{dt}=K\mathbf{z}(t),
\end{equation*}
where $K\in\mathbb{R}^{m\times m}$ is the Koopman matrix.
This yields the formal solution
\begin{equation*}
    \mathbf{z}(t)=\mathrm{e}^{Kt}\mathbf{z}(0).
\end{equation*}
The linear latent representation provides a structured reduced-order description of the epidemic dynamics and enables stable forecasting beyond the observation interval.

\subsection{State Reconstruction}\label{S3.3}
To recover the epidemic variables from the lifted coordinates, we introduce a decoder network
\begin{equation*}
   \hat{\mathbf{x}}(t)=\Psi_{\omega}(\mathbf{z}(t)),
\end{equation*}
where $\Psi_{\omega}\colon\mathbb{R}^m\to\mathbb{R}^5$ is a neural network with parameters $\omega$.
The full lifted reconstruction is therefore
\begin{equation*}
   \hat{\mathbf{x}}(t)=\Psi_{\omega}\bigl(\Phi_{\theta}(\mathbf{x}(t))\bigr).
\end{equation*}
The decoder maps the latent observables back to physically meaningful epidemic compartments.

\subsection{Physics-Informed Residuals}\label{S3.4}
Let
\begin{equation*}
   \hat{\mathbf{x}}(t) = \bigl[\hat{s}(t), \,\hat{e}(t), \,\hat{i}(t), \,\hat{r}(t), \,\hat{d}(t)\bigr]^\top,
\end{equation*}
denote the reconstructed solution. 
The normalized SEIRSD residual operator is defined by
\begin{equation*}
\begin{aligned}
\mathcal{R}(t) := \mathcal{R}(\hat{\mathbf{x}}(t)) &= \big[\mathcal{R}_s(t), \,\mathcal{R}_e(t), \,\mathcal{R}_i(t), \,\mathcal{R}_r(t), \,\mathcal{R}_d(t)\big]^{\top},\\
&= \frac{d\hat{\mathbf{x}}}{dt} - F(\hat{\mathbf{x}}),
\end{aligned}
\end{equation*}
where $F$ denotes the right hand side of the system \eqref{E2.3}.
These residuals are computed using automatic differentiation and are used to penalize deviations from the governing SEIRSD equations during training.

\subsection{Composite Loss Function}\label{S3.5}
The K--PINN training objective is given by
\begin{equation*}
    \mathcal{L} = \lambda_d\mathcal{L}_{\rm data} + \lambda_k\mathcal{L}_{\rm koop} + \lambda_p\mathcal{L}_{\rm phys} + \lambda_c\mathcal{L}_{\rm cons},
\end{equation*}
where $\lambda_d,\lambda_k,\lambda_p,\lambda_c>0$.
Let $N_d$ be the number of observed epidemic snapshots, and $N_r$ be the number of collocation points used to evaluate the residuals of the governing equations. Thus,
\begin{itemize}
\item The data reconstruction loss is given by
\begin{equation*}
\mathcal{L}_{\rm data} = \frac{1}{N_d}\sum_{j=1}^{N_d} \bigl\| \mathbf{x}_j-\widehat{\mathbf{x}}_j \bigr\|_2^2.
\end{equation*}

\item The Koopman consistency loss is given by
\begin{equation*}
\mathcal{L}_{\rm koop} = \frac{1}{N_d}\sum_{j=1}^{N_d} \Bigl\| \frac{d\mathbf{z}_j}{dt}-K\mathbf{z}_j \Bigr\|_2^2.
\end{equation*}

\item The physics loss is given by
\begin{equation*}
\mathcal{L}_{\rm phys} = \frac{1}{N_r}\sum_{j=1}^{N_r} \bigl\| \mathcal{R}(t_j) \bigr\|_2^2.
\end{equation*}

\item The constraint term is given by
\begin{equation*}
\mathcal{L}_{\rm cons} = \mathcal{L}_{\rm pos} + \mathcal{L}_{\rm param},
\end{equation*}
where $\mathcal{L}_{\rm pos}$ penalizes negative population values and $\mathcal{L}_{\rm param}$ regularizes the epidemiological parameters.
\end{itemize}

\subsection{Epidemiological Constraints}\label{S3.6}
Since the state variables are normalized population fractions, they satisfy
\begin{equation*}
0 \le s(t), e(t), i(t), r(t), d(t) \le 1 \ \text{ and } \ s(t) + e(t) + i(t) + r(t) + d(t) = 1.
\end{equation*}
To ensure admissible parameter values, we can parameterize
\begin{equation*}
    \beta = \exp(\tilde{\beta}), \ \delta = \exp(\tilde{\delta}), \ \gamma = \exp(\tilde{\gamma}), \ \text{ and } \ \mu = \exp(\tilde{\mu}),
\end{equation*}
with unconstrained variables $\tilde{\beta}, \tilde{\delta}, \tilde{\gamma}, \tilde{\mu} \in \mathbb{R}$.

\subsection{Inverse Learning Problem}\label{S3.7}
Let the full set of trainable parameters be
\begin{equation*}
\Theta = \{\theta, \,\omega, \,K, \,\beta, \,\delta, \,\gamma, \,\mu\}.
\end{equation*}
The inverse problem is formulated as
\begin{equation*}
\Theta^\star = \arg\min_{\Theta}\mathcal{L}(\Theta).
\end{equation*}
This single optimization simultaneously learns the observable map, the Koopman operator, the decoder, and the epidemiological parameters.

\section{Numerical Methodology}\label{S4}
The numerical experiments are designed to evaluate the K--PINN proposed for the system under different observation regimes. 
The main objectives are to reconstruct epidemic trajectories, estimate unknown parameters, and forecast future dynamics based on sparse or noisy observations. 
Depending on the experiment, the model may use full-state data, partial observations, or perturbed measurements.
To ensure consistency between the mathematical model, the learning framework, and the numerical experiments, the same normalized formulation is used throughout.

\subsection{Synthetic Data Generation via a Nonstandard Finite Difference Scheme}\label{S4.1}
Synthetic trajectories are generated using an NSFD scheme, which is particularly suitable for epidemic models because it preserves positivity, boundedness, structural consistency, and dynamical consistency. 
Unlike standard time discretizations, which may introduce nonphysical behavior such as negative compartment values, NSFD methods are designed to avoid such artifacts.

Let $\{t_0,t_1,\dots,t_M\}$ be a uniform time grid with a step size $k = \Delta t$. 
In the NSFD framework, the derivative of a generic variable $u(t)$ is approximated by
\begin{equation*}
    \dot{u}(t_n)\approx \frac{u^{n+1}-u^n}{\varphi(k)},
\end{equation*}
where $\varphi(k)$ is a nontrivial denominator function that satisfies
\begin{equation*}
    \varphi(k)>0, \ \text{ and } \ \varphi(k)=k+\mathcal{O}(k^2).
\end{equation*}
A common choice is
\begin{equation*}
    \varphi(k)=\frac{e^{\eta k}-1}{\eta},
\end{equation*}
where $\eta$ is a stabilization parameter obtained by  comparing with the asymptotic behavior of the total population \cite{maamar2024nonstandard}.
In the absence of natural mortality, one may simply set $\eta=0$,
yielding $\varphi(k)=k$.

Let $s^n\approx s(t_n)$, $e^n\approx e(t_n)$, $i^n\approx i(t_n)$, $r^n\approx r(t_n)$, and $d^n\approx d(t_n)$. The NSFD discretization of the SEIRSD system~\eqref{E2.3} is written as
\begin{equation*}
\left\{\begin{aligned}
\frac{s^{n+1}-s^n}{\varphi(k)} &= -\beta \frac{s^{n+1}i^n}{1-d^n}+\alpha r^n,\\
\frac{e^{n+1}-e^n}{\varphi(k)} &= \beta \frac{s^{n+1}i^n}{1-d^n}-\sigma e^{n+1},\\
\frac{i^{n+1}-i^n}{\varphi(k)} &= \sigma e^{n+1}-(\gamma+\mu)i^{n+1},\\
\frac{r^{n+1}-r^n}{\varphi(k)} &= \gamma i^{n+1}-\alpha r^{n+1}.
\end{aligned}\right.
\end{equation*}
Afterwards,
\begin{equation}\label{E4.1}
\left\{\begin{aligned}
 s^{n+1} &= \frac{s^n + \alpha \varphi(k) r^n}{1 + \beta \varphi(k) \frac{i^n}{1-d^n}},\\
 e^{n+1} &= \frac{e^n + \beta \varphi(k) \frac{s^{n+1} i^n}{1-d^n}}{1 + \sigma \varphi(k)},\\
 i^{n+1} &= \frac{i^n + \sigma \varphi(k) e^{n+1}}{1 + (\gamma+\mu) \varphi(k)},\\
 r^{n+1} &= \frac{r^n + \gamma \varphi(k) i^{n+1}}{1 + \alpha \varphi(k)},
\end{aligned}\right.
\end{equation}
with the deceased proportion updated consistently by
\begin{equation}\label{E4.2}
   d^{n+1}=1-s^{n+1}-e^{n+1}-i^{n+1}-r^{n+1}.
\end{equation}

This discretization adheres to standard NSFD design principles. 
First, the nonlinear incidence term $\beta \frac{s i}{1-d}$ is treated semi-implicitly by evaluating $s$ at the new time level and $i$ at the current time level. This preserves positivity while ensuring that the scheme is solvable sequentially. 
Second, the recovery and progression terms are written to maintain the correct flow between compartments. 
Third, the update for $d^{n+1}$ is obtained from the population conservation relation, ensuring that the discrete total population remains equal to one.
For more details about the NSFD analysis, we encourage readers to consult \cite{Ehrhardt2013, Zinihi2025NSFD} and the references therein.

The scheme is designed to preserve the nonnegativity of all compartments, keeping the solution inside the invariant simplex $\Delta$. 
This makes NSFD trajectories suitable for use as reliable synthetic data for training the proposed K--PINN and for validating its reconstruction and forecasting performance. 
In the numerical experiments, the time step $k$ is chosen small enough to accurately capture the epidemic dynamics accurately while maintaining computational efficiency. 
Due to the limited availability of real data, we set $\alpha = 0$ in Section~\ref{S5}.

\subsection{Data Collection and Koopman Matrix Estimation}\label{S4.2}
The K--PINN framework requires state trajectory data. 
This data can be obtained either from:
numerical simulations (synthetic data) or empirical epidemiological time series, after appropriate normalization.
We discretize the trajectory at uniform sampling times $t_n = nk = n \Delta t$, with $n = 0, 1, \dots ,M$
This produces a discrete sequence of state snapshots
\begin{equation*}
\mathbf{x}_n = \big[s(t_n), \,e(t_n), \,i(t_n), \,r(t_n), \,d(t_n)\big]^\top.
\end{equation*}
To construct the Koopman representation, each snapshot is mapped into an observable space through a selected set of functions
\begin{equation*}
\mathbf{y}_n = \big[\psi_1(\mathbf{x}_n), \,\psi_2(\mathbf{x}_n), \,\ldots, \,\psi_m(\mathbf{x}_n)\big]^\top,
\end{equation*}
where $\{\psi_j\}_{j=1}^{m}$ denotes the proposed observable dictionary.
In the present work, this dictionary may be chosen to reflect the epidemic structure by including the compartment variables and selected nonlinear interaction terms that capture infection and coupling effects, as described in Section~\ref{S3.1}.
The lifted snapshot matrices are then defined as
\begin{equation*}
    Y = \bigl(\mathbf{y}_0,\mathbf{y}_1,\ldots,\mathbf{y}_{M-1}\bigr)\in\mathbb{R}^{m\times M}, \ \text{ and } \  Y' = \bigl(\mathbf{y}_1,\mathbf{y}_2,\ldots,\mathbf{y}_M\bigr)\in\mathbb{R}^{m\times M}.
\end{equation*}
In the lifted space, the Koopman evolution is approximated by
\begin{equation*}
\mathbf{y}_{k+1}\approx K\mathbf{y}_k,
\end{equation*}
where $K\in\mathbb{R}^{m\times m}$ is the finite-dimensional Koopman matrix. 
Consequently, the data matrices satisfy
\begin{equation*}
Y'\approx KY.
\end{equation*}
The Koopman matrix is identified by the least-squares problem
\begin{equation*}
K=\arg\min_{\widetilde K\in\mathbb{R}^{m\times m}}
\bigl\|Y'-\widetilde K Y\bigr\|_F,
\end{equation*}
whose closed-form solution is given by
\begin{equation*}
K=Y'Y^\star,
\end{equation*}
where $Y^\star$ denotes the Moore-Penrose pseudoinverse of $Y$.

The eigen-decomposition of $K$ provides approximations of the Koopman eigenvalues and modes, which describe the dominant growth, decay, and oscillatory patterns of the epidemic dynamics in the lifted space. 
These quantities are useful for analyzing the reduced-order structure of the system and for comparing the learned latent dynamics with the proposed K--PINN representation. 
After computing the Koopman matrix, one can propagate future trajectories in the observable space. This provides an additional numerical benchmark for the model's forecasting performance.

\subsection{Training Strategy and Forecasting Procedure}\label{S4.3}
The proposed K--PINN is trained by minimizing the composite loss defined in Section 3. 
In practice, training begins with a short pretraining phase based on the reconstruction term. This phase  helps initialize the encoder, decoder, and latent Koopman operator in a stable regime. 
Following this initialization stage, the full objective is optimized to enforce data fidelity, physics residuals, Koopman consistency, and admissibility constraints simultaneously.

The training data consist of two sets of points. 
The first set contains the observed epidemic snapshots used for reconstruction loss. The second set contains collocation points sampled across the temporal domain for evaluating the residuals. 
Then, automatic differentiation is used to compute the time derivatives of the network outputs, which allows the physics-informed terms to be evaluated without discretization error.

Adam is used for optimization in the early stage because it is robust to noisy gradients and helps the network explore the parameter space. 
Once the loss has stabilized, L-BFGSS\footnote{A memory-efficient quasi-Newton optimization method used to refine neural network training and accelerate convergence near the optimum}
is applied to refine the solution and improve convergence accuracy.
This two-stage strategy is effective for PINN-type models, especially when the latent Koopman matrix and epidemiological parameters are learned simultaneously.

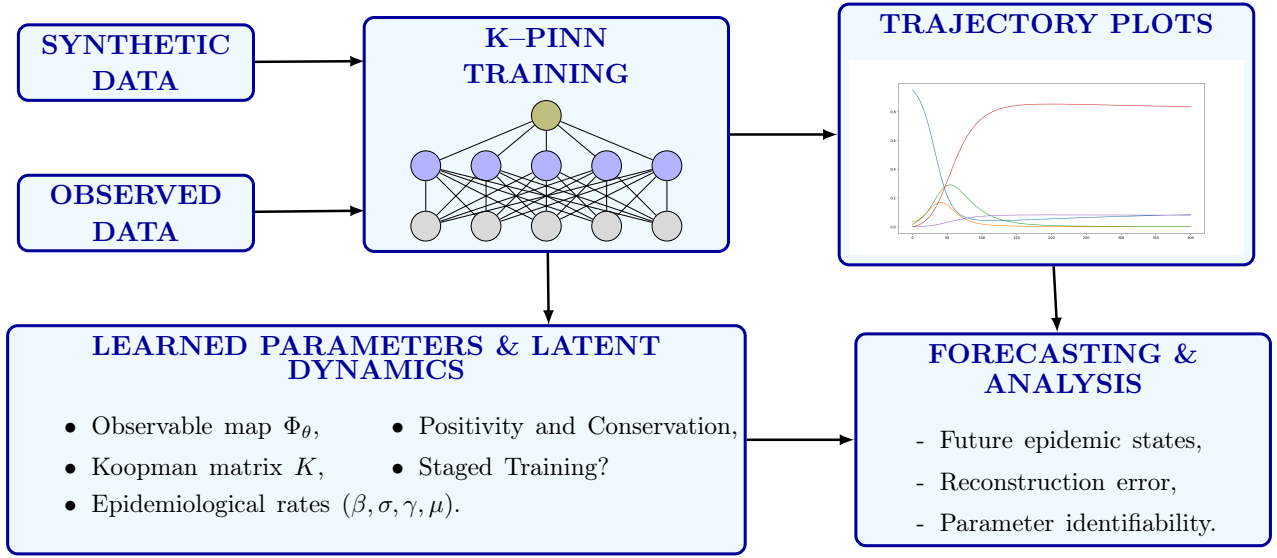
\begin{figure}[H]
\centering
\adjustbox{max width=\textwidth}{
\begin{tikzpicture}[node distance=3cm]
\node (SD) [draw=blue!60!black, very thick, rounded corners,
font=\bfseries, text width=3cm, fill=cyan!5, align=center] 
{\small\centering{\color{blue!60!black}SYNTHETIC DATA}};
\node (RD) [draw=blue!60!black, very thick, rounded corners,
font=\bfseries, text width=3cm, fill=cyan!5, align=center, below = of SD, yshift=2cm] 
{\small\centering{\color{blue!60!black}OBSERVED\\ DATA}};
\node (Training) [draw=blue!60!black, very thick, rounded corners,
font=\bfseries, text width=4.8cm, fill=cyan!5, align=center, right = of SD, yshift=-1cm, xshift=-1.5cm] 
{\small\centering{\color{blue!60!black}K--PINN\\ TRAINING\vspace{0.2cm}}
\begin{tikzpicture}[scale=0.7]
\node[circle, draw, fill=green!50!red!50, minimum size=0.4cm] (i0) at (2.4, 0) {};
\node[circle, draw, fill=blue!30, minimum size=0.4cm] (i1) at (0, -1) {};
\node[circle, draw, fill=blue!30, minimum size=0.4cm] (i2) at (1.2, -1) {};
\node[circle, draw, fill=blue!30, minimum size=0.4cm] (i3) at (2.4, -1) {};
\node[circle, draw, fill=blue!30, minimum size=0.4cm] (i4) at (3.6, -1) {};
\node[circle, draw, fill=blue!30, minimum size=0.4cm] (i5) at (4.8, -1) {};
\node[circle, draw, fill=gray!30, minimum size=0.4cm] (i6) at (0, -2.2) {};
\node[circle, draw, fill=gray!30, minimum size=0.4cm] (i7) at (1.2, -2.2) {};
\node[circle, draw, fill=gray!30, minimum size=0.4cm] (i8) at (2.4, -2.2) {};
\node[circle, draw, fill=gray!30, minimum size=0.4cm] (i9) at (3.6, -2.2) {};
\node[circle, draw, fill=gray!30, minimum size=0.4cm] (i10) at (4.8, -2.2) {};
\foreach \i in {i1,i2,i3,i4,i5}
\foreach \j in {i0}
\draw (\i) -- (\j);
\foreach \i in {i1,i2,i3,i4,i5}
\foreach \h in {i6,i7,i8,i9,i10}
\draw (\i) -- (\h);
\end{tikzpicture}
};
\node (Plots) [draw=blue!60!black, very thick, rounded corners,
font=\bfseries, text width = 5.5cm, fill=cyan!5, right = of Training, xshift=-1.5cm] 
{\small\centering{\color{blue!60!black}TRAJECTORY PLOTS} \footnotesize
\begin{equation*}
\includegraphics[width=1\textwidth]{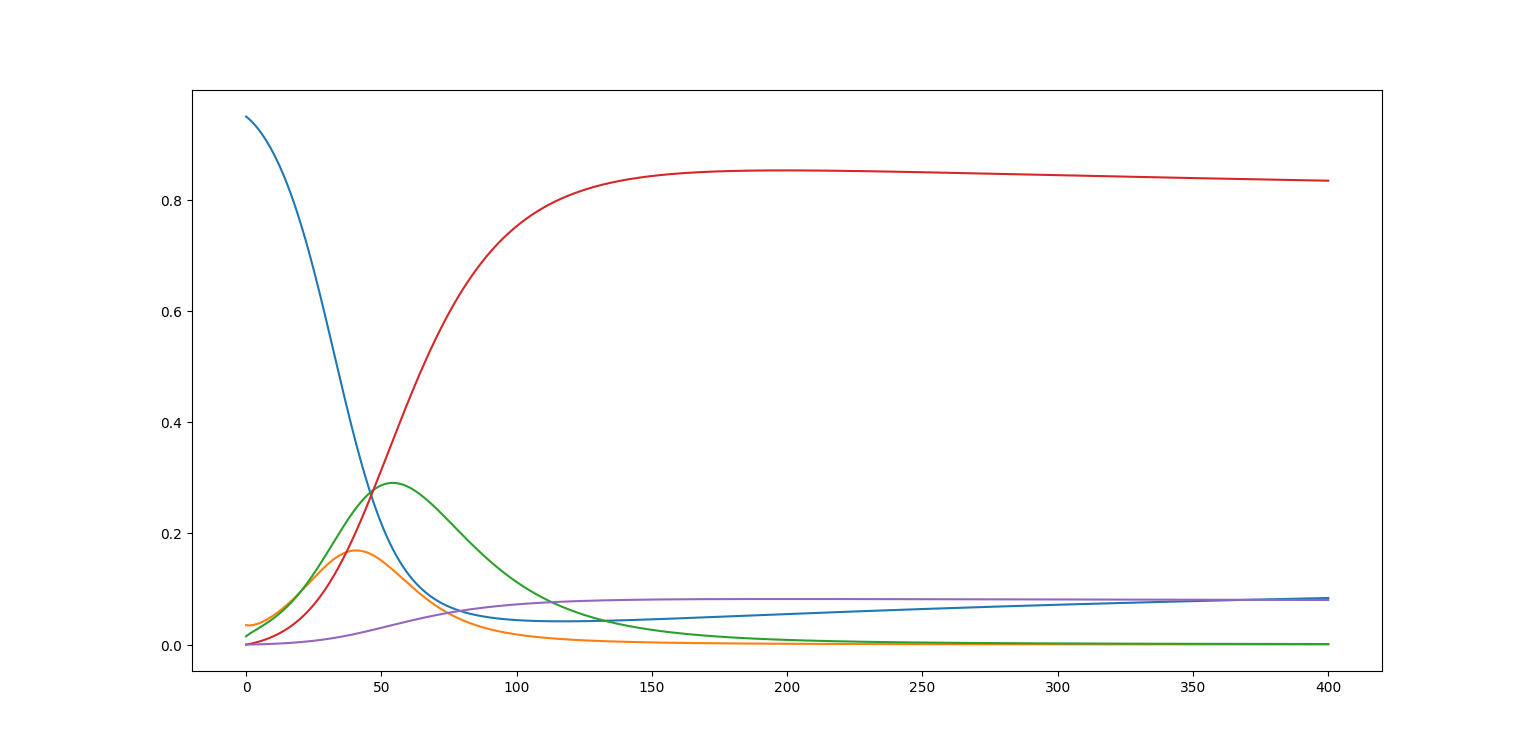}
\end{equation*}
};
\node (Learning) [draw=blue!60!black, very thick, rounded corners,
font=\bfseries, text width=10cm, fill=cyan!5, align=center, below = of RD, yshift=2cm, xshift=3.35cm] 
{\small\centering{\color{blue!60!black}LEARNED PARAMETERS \& LATENT
DYNAMICS}\scriptsize

\raggedright{
\begin{table}[H]
\begin{minipage}{0.44\linewidth}
\begin{itemize}\footnotesize
\item Observable map $\Phi_\theta$,
\vspace{-0.2cm}
\item Koopman matrix $K$,
\end{itemize}
\end{minipage}
\hfill
\begin{minipage}{0.63\linewidth}
\begin{itemize}\footnotesize
\item Positivity and Conservation,
\vspace{-0.2cm}
\item Staged Training?
\end{itemize}
\end{minipage}
\begin{itemize}\footnotesize
\vspace{-0.2cm}
\item Epidemiological rates $(\beta,\sigma,\gamma,\mu)$.
\end{itemize}
\end{table}
}};
\node (Unc) [draw=blue!60!black, very thick, rounded corners,
font=\bfseries, text width = 5.5cm, fill=cyan!5, right = of Learning, xshift=-1.5cm] 
{\small\centering{\color{blue!60!black}FORECASTING \&
ANALYSIS} \footnotesize

\raggedright{
\begin{table}[H]
\begin{minipage}{\linewidth}
\begin{itemize}\footnotesize
\item[-] {\footnotesize Future epidemic states,}
\vspace{-0.2cm}
\item[-] Reconstruction error,
\vspace{-0.2cm}
\item[-] Parameter identifiability.
\end{itemize}
\end{minipage}
\end{table}
}};
\draw [->, thick, >=latex, line width=1pt] (SD) -- (3.19,0.02);
\draw [->, thick, >=latex, line width=1pt] (RD) -- (3.19,-2.06);
\draw [->, thick, >=latex, line width=1pt] (Training) -- (5.75,-3.65);
\draw [->, thick, >=latex, line width=1pt] (Training) -- (Plots);
\draw [->, thick, >=latex, line width=1pt] (Plots) -- (Unc);
\draw [->, thick, >=latex, line width=1pt] (Learning) -- (Unc);
\end{tikzpicture}
}
\captionof{figure}{Workflow of the proposed K--PINN framework for epidemic reconstruction and forecasting.}\label{F2}
\end{figure}

After training, the learned latent dynamics are used for forecasting. 
Starting from the encoded initial condition, the Koopman operator propagates the observables forward in time, and the decoder reconstructs the epidemic compartments at future times. 
This produces a structured forecasting mechanism that combines linear latent evolution with nonlinear epidemic constraints. 
Figure~\ref{F2} summarizes the complete K--PINN pipeline, from NSFD data generation and observable lifting to joint optimization and forecasting. 

Algorithm \ref{Alg-KPINN} outlines the step-by-step training and prediction procedure used in the numerical experiments.
\begin{algorithm}[H]
\caption{Training and Forecasting Procedure for the Proposed K--PINN}\label{Alg-KPINN}
\begin{algorithmic}[1]
\STATE \emph{Input:} Epidemic trajectories, observed snapshots, collocation points, initial parameters.
\STATE \emph{Initialization:} Initialize the encoder $\Phi_{\theta}$, decoder $\Psi_{\omega}$, Koopman matrix $K$, and epidemiological parameters $(\beta, \,\sigma, \,\gamma, \,\mu)$.
\STATE \emph{Data preparation:} Sample observation points $\{t_j\}_{j=1}^{N_d}$ and collocation points $\{ \tau_j \}_{j=1}^{N_r}$ over the time interval.
\STATE \emph{Pretraining stage:} Minimize the reconstruction loss $\mathcal{L}_{\rm data}$ using Adam to stabilize the encoder and decoder.
\STATE \emph{Joint training stage:} Minimize the full composite loss
\begin{equation*}
\mathcal{L}=\lambda_d\mathcal{L}_{\rm data}
+\lambda_k\mathcal{L}_{\rm koop}
+\lambda_p\mathcal{L}_{\rm phys}
+\lambda_c\mathcal{L}_{\rm cons}
\end{equation*}
using Adam.
\STATE \emph{Physics evaluation:} Use automatic differentiation to compute the residuals of the normalized system at collocation points.
\STATE \emph{Koopman update:} Update the latent operator $K$ jointly with the neural parameters.
\STATE \emph{Refinement stage:} Switch to L-BFGS and continue optimization until convergence.
\STATE \emph{Forecasting:} Propagate the latent state forward using $\mathbf{z}_{k+1}=K\mathbf{z}_k$, then reconstruct future epidemic states using $\hat{\mathbf{x}}_{k+1}=\Psi_{\omega}(\mathbf{z}_{k+1})$.
\STATE \emph{Output:} Reconstructed trajectories, estimated parameters, and future predictions.
\end{algorithmic}
\end{algorithm}

\subsection{Error Metrics}\label{S4.4}
Several standard error metrics are employed to evaluate the accuracy of the proposed framework.
For an individual compartment, e.g.\ the susceptible population, 
the root mean square error (RMSE) is defined as follows
\begin{equation*}
   \mathrm{RMSE}(s) = \sqrt{\frac{1}{M+1}\sum_{n=0}^{M}
    \bigl(s_n-\hat{s}_n\bigr)^2},
\end{equation*}
while the mean absolute error (MAE) is given by
\begin{equation*}
   \mathrm{MAE}(s) = \frac{1}{M+1}\sum_{n=0}^{M}
   \bigl|s_n-\hat{s}_n\bigr|.
\end{equation*}
In addition, the relative error is computed as
\begin{equation*}
    \mathrm{RelErr}(s) = \frac{\|s-\hat{s}\|_2} {\|s\|_2} \times 100\%.
\end{equation*}
Similar expressions are used for the remaining compartments.
To evaluate the reconstruction accuracy of the entire epidemic state, we consider the discrete $L^2$ error
\begin{equation*}
    E_{L^2} = \sqrt{\frac{1}{M+1} \sum_{n=0}^{M} \|\mathbf{x}_n-\hat{\mathbf{x}}_n\|_2^2},
\end{equation*}
and the maximum error
\begin{equation*}
    E_{L^\infty} = \max_{0\le n\le M} \|\mathbf{x}_n-\hat{\mathbf{x}}_n\|_\infty.
\end{equation*}
These metrics are used to compare the K--PINN with the baseline methods in the numerical results section.

\section{Numerical Results}\label{S5}
We conducted numerical experiments on both synthetic monkeypox (Mpox) data and real-world SARS-CoV-2 (Covid-19) epidemic data to evaluate the proposed K--PINN framework. 
Throughout the simulations, we set $\alpha = 0$.

First, we consider synthetic epidemic trajectories generated to emulate Mpox dynamics, where the true parameters $(\beta, \sigma, \gamma, \mu)$ are known. 
This allows us to rigorously assess the framework's ability to recover hidden epidemiological parameters from partial and noisy observations using the training procedure described in Section~\ref{S4}.
We also compare the K--PINN approach with classical PINNs and K--EDMD methods to benchmark its performance in terms of identifiability and stability of the learned dynamics.
After validating the methodology on synthetic data, we apply it to a real-world SARS-CoV-2 dataset from Sweden, focusing on confirmed, recovered, and deceased cases.

Section~\ref{S5.1} presents results for the synthetic experiments, where data are generated using the NSFD scheme~\eqref{E4.1}--\eqref{E4.2}. 
Section~\ref{S5.2} reports the results on real data, emphasizing predictive accuracy, parameter inference, and epidemiological interpretability.

\subsection{Synthetic Mpox Data Experiments}\label{S5.1}
The epidemiological parameters were selected within biologically plausible ranges reported in recent Mpox studies, as summarized in Table~\ref{Tab1}, particularly for clade Ib outbreaks during 2024--s2025.
We adopted representative values within these intervals to generate the data: 
 $\beta = 0.25$, $\sigma = 0.13$, $\gamma = 0.052$, and $\mu = 0.005$.

\begin{table}[H]
\centering
\setlength{\tabcolsep}{0.5cm}
\caption{Epidemiological parameters for the proposed model applied to Mpox (clade Ib, 2024–2025).}\label{Tab1}
\adjustbox{max width=\textwidth}{
\begin{tabular}{cccccc}
\hline
\textbf{Epidemic} & $\beta$ & $\sigma$ & $\gamma$ & $\mu$ & \textbf{References} \\
\hline\hline
Mpox & 0.1 – 0.3 & 0.077 – 0.2 & 0.036 – 0.071 & 0.001 – 0.03 & \cite{WHO2024Mpox, Lancet2024} \\
\hline
\end{tabular}
}
\end{table}

\begin{figure}[H]
\centering
\includegraphics[width=1\textwidth]{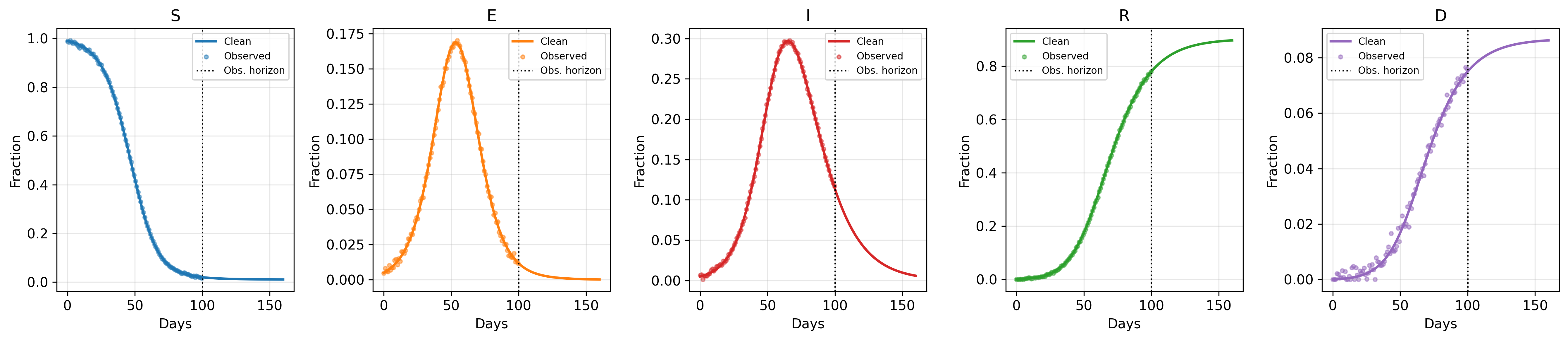}
\caption{Synthetic epidemic trajectories generated using the NSFD scheme~\eqref{E4.1}--\eqref{E4.2}.}\label{F3}
\end{figure}

Figure~\ref{F3} presents the synthetic trajectories of the susceptible, exposed, infected, recovered, and deceased compartments over the considered time horizon. 
These trajectories serve as the ground-truth dataset used throughout the numerical experiments.

\begin{figure}[H]
\centering
\includegraphics[width=1\textwidth]{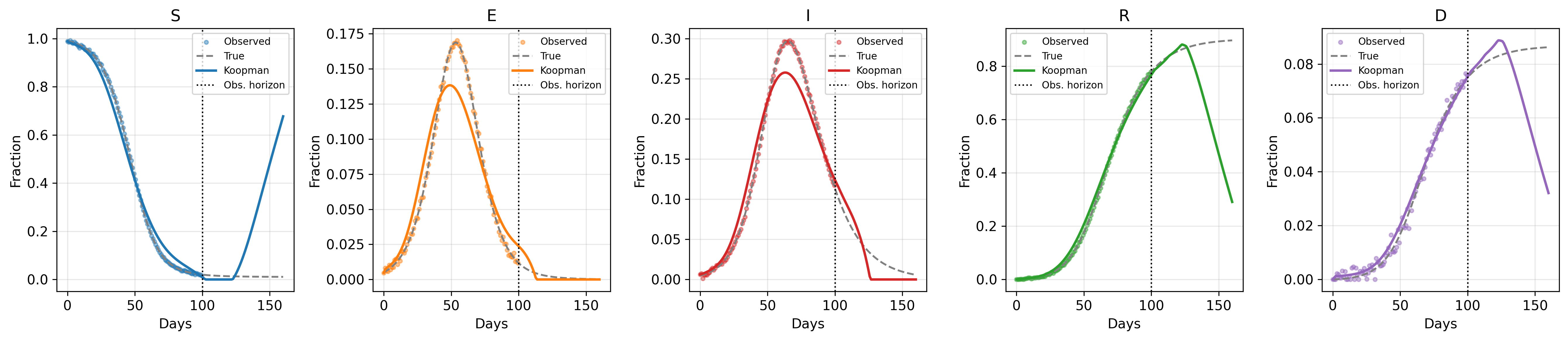}
\caption{State reconstruction and forecasting using the K--EDMD framework for the synthetic dataset.}\label{F4}
\end{figure}

As a first benchmark, the epidemic dynamics were approximated using the classical EDMD approach. 
The observable dictionary introduced in Section~\ref{S3.1} was employed to create a finite-dimensional approximation of the Koopman operator. 
Figure~\ref{F4} compares the reconstructed trajectories and future forecasts obtained through K--EDMD with the reference synthetic data. 
Although K--EDMD captures the overall qualitative behavior of the epidemic evolution, discrepancies become more pronounced during long-term forecasting due to the lack of explicit enforcement of the governing epidemic equations.
Diagnostic quantities associated with this approach are presented in Figure~\ref{FA2}.

\begin{figure}[H]
\centering
\includegraphics[width=1\textwidth]{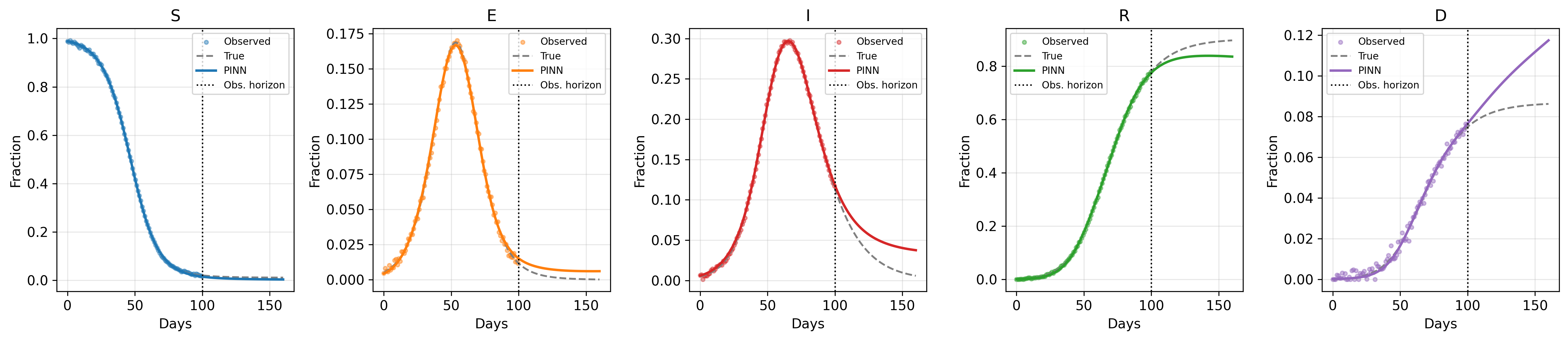}
\caption{Epidemic state reconstruction and forecasting using the PINN framework for the synthetic dataset.}\label{F5}
\end{figure}

Next, we consider a standard PINN implementation in which the normalized equations are enforced via a  physics-based loss function, but Koopman latent dynamics are not incorporated. 
Figure~\ref{F5} illustrates the reconstructed epidemic trajectories and forecasting results. 
The PINN framework successfully reconstructs the observed trajectories and provides accurate estimates of the epidemiological parameters, as shown in Figures~\ref{FA3} and \ref{FA4}.
However, the forecasting performance may deteriorate over extended prediction horizons due to the difficulty of learning strongly nonlinear epidemic dynamics directly in the original state space.

\begin{figure}[H]
\centering
\includegraphics[width=1\textwidth]{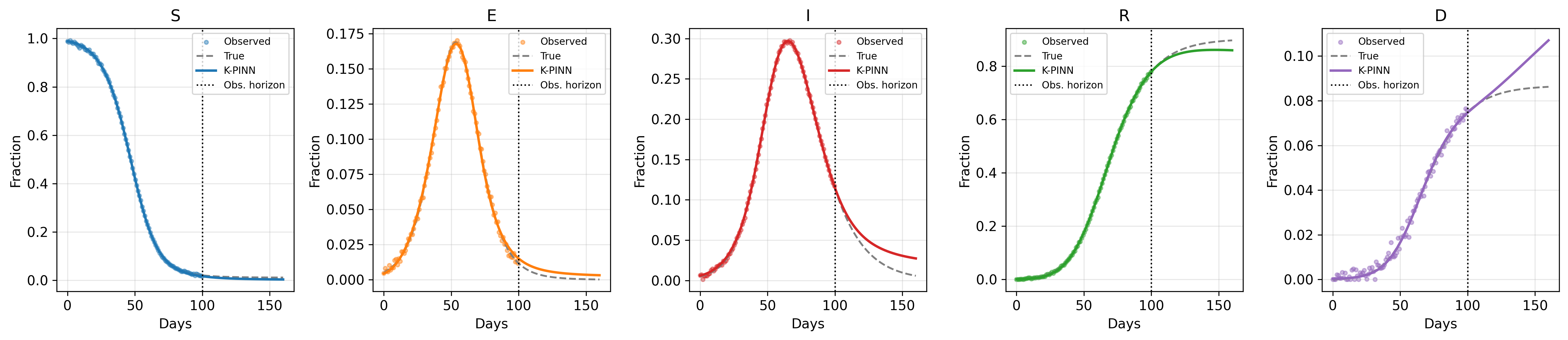}
\caption{Epidemic state reconstruction and forecasting using the proposed K--PINN framework for the synthetic dataset.}\label{F6}
\end{figure}

\begin{figure}[H]
\centering
\includegraphics[width=1\textwidth]{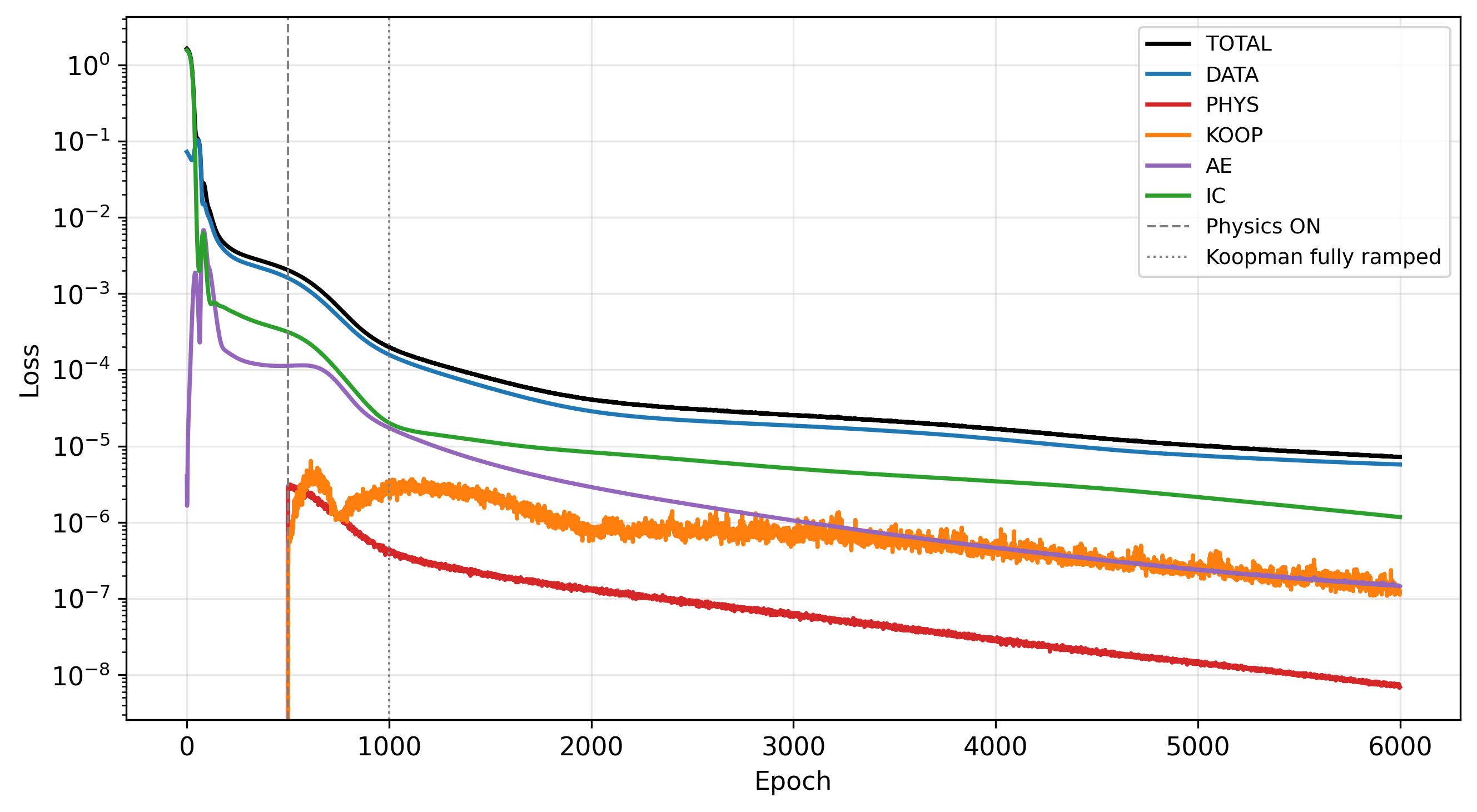}
\caption{Evolution of the composite K--PINN loss function during training for the synthetic dataset.}\label{F7}
\end{figure}

\begin{figure}[H]
\centering
\includegraphics[width=1\textwidth]{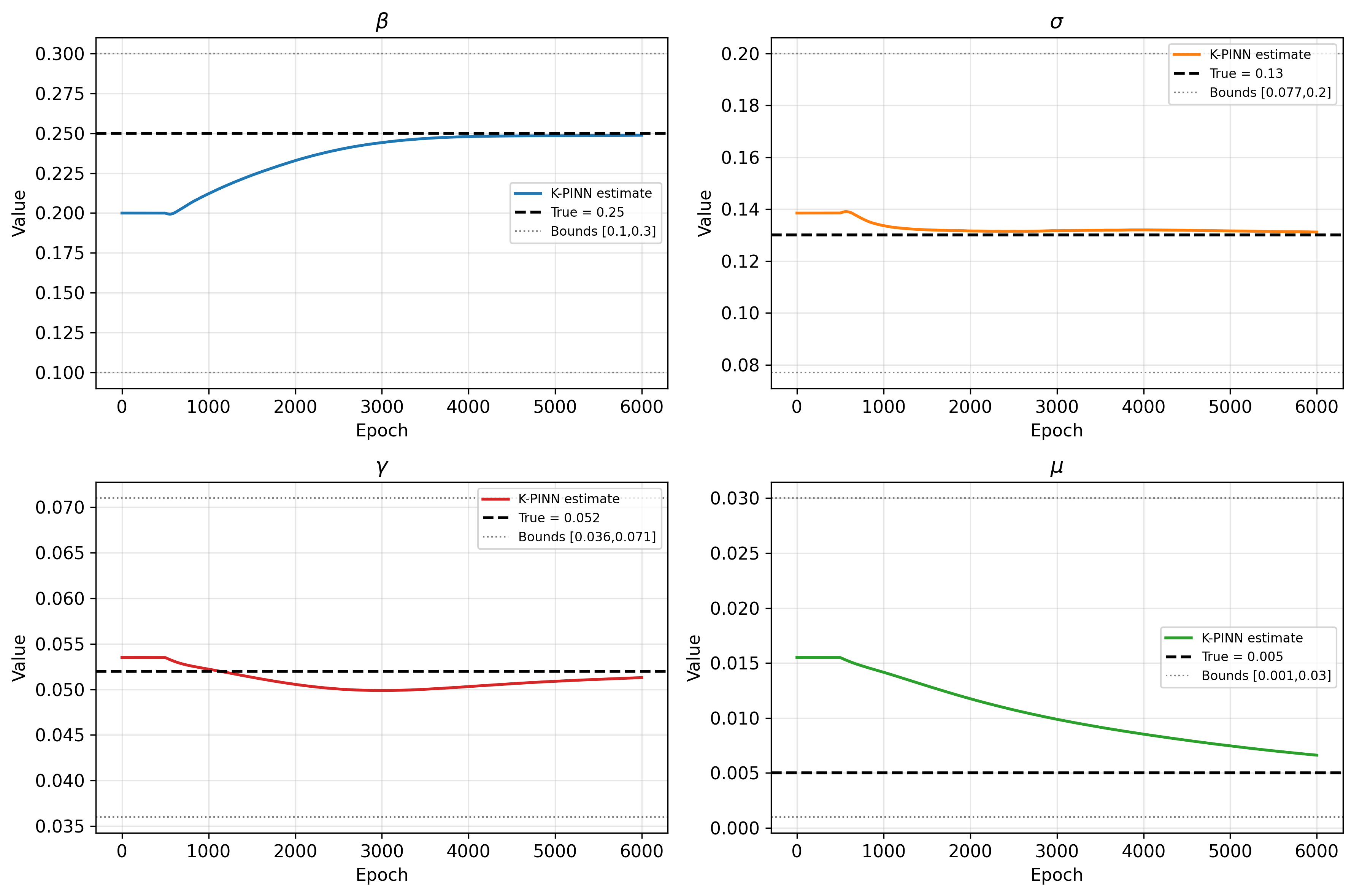}
\caption{Convergence of epidemiological parameter estimates obtained by the proposed K--PINN framework for the synthetic dataset.}\label{F8}
\end{figure}

Finally, the proposed K--PINN framework is applied to the same synthetic dataset. 
Figure~\ref{F6} presents the reconstructed epidemic trajectories together with long-horizon forecasts. 
Figure~\ref{F7} displays the convergence behavior of the composite loss function, demonstrating the stable optimization of the data, physics, Koopman, and constraint terms. 
Figure~\ref{F8} illustrates the evolution of the estimated epidemiological parameters throughout training. 
The results show that K--PINN accurately reconstructs all epidemic compartments while simultaneously providing reliable parameter estimates and improved forecasting capability. 
The incorporation of Koopman latent dynamics significantly enhances stability and predictive performance compared with both the standalone K--EDMD and PINN approaches.

For clarity of presentation, Figure~\ref{FA1} displays the collective reconstruction and forecasting results obtained with the proposed K--PINN framework. 
Detailed, compartment-wise comparisons for all epidemic variables and all competing approaches are provided in Appendix~\ref{App1}.
Table~\ref{Tab2} reports the reconstruction and forecasting errors obtained by the three competing approaches, providing a quantitative assessment of the proposed methodology. 
Table~\ref{Tab3} summarizes the parameter estimation results and compares the recovered epidemiological parameters with their reference values.
\begin{table}[H]
\centering
\setlength{\tabcolsep}{0.3cm}
\caption{Comparison of reconstruction and forecasting errors obtained using K--EDMD, PINN, and the proposed K--PINN framework for the synthetic Mpox dataset.}\label{Tab2}
\adjustbox{max width=\textwidth}{
\begin{tabular}{cccccccc}
\hline
\textbf{Method} & $E_{L^2}$ & $E_{L^\infty}$ & RMSE$(S)$ & RMSE$(E)$ & RMSE$(I)$ & RMSE$(R)$ & RMSE$(D)$ \\
\hline\hline
K--EDMD & 0.24420 & 0.66615 & 0.18123 & 0.01317 & 0.01939 & 0.16138 & 0.01398 \\
PINN & 0.03130 & 0.06097 & 0.00456 & 0.00343 & 0.01468 & 0.02469 & 0.01106 \\
K--PINN & 0.01839 & 0.03722 & 0.00422 & 0.00242 & 0.00966 & 0.01353 & 0.00618 \\
\hline
\end{tabular}
}
\end{table}
\begin{table}[H]
\centering
\setlength{\tabcolsep}{0.5cm}
\caption{True and estimated epidemiological parameters for the synthetic SEIRSD experiment.}\label{Tab3}
\begin{tabular}{ccccc}
\hline
\textbf{Parameters} & $\beta$ & $\sigma$ & $\gamma$ & $\mu$\\
\hline\hline
True value & 0.2500 & 0.1300 & 0.0520 & 0.0050 \\
\hline
PINN estimate & 0.2479 & 0.1312 & 0.0512 & 0.0067 \\
\hline
K--PINN estimate & 0.2488 & 0.1310 & 0.0513 & 0.0066 \\
\hline
\end{tabular}
\end{table}

\subsection{Real-World COVID-19 Data Experiments}\label{S5.2}
To further evaluate the applicability of the proposed K--PINN framework, we examine real-world epidemic data from Germany, Morocco, and Sweden during the COVID-19 outbreaks. We selected these countries because they have reliable daily case and mortality reports and distinct epidemic dynamics. 
The corresponding epidemiological parameters are constrained within biologically plausible ranges, as summarized in Table~\ref{Tab4}.
\begin{table}[H]
\centering
\setlength{\tabcolsep}{0.5cm}
\caption{Epidemiological parameters for the proposed model applied to COVID-19 in Europe (2024).}\label{Tab4}
\adjustbox{max width=\textwidth}{
\begin{tabular}{cccccc}
\hline
\textbf{Epidemic} & $\beta$ & $\sigma$ & $\gamma$ & $\mu$ & \textbf{References} \\
\hline\hline
COVID-19 & 0.2 – 0.4 & 0.1 – 0.3 & 0.05 – 0.1 & 0.001 – 0.01 & \cite{UKGov2025, WHO2024Covid} \\
\hline
\end{tabular}
}
\end{table}
The daily number of confirmed cases and deaths comes from publicly available epidemiological databases \cite{Gehrcke2025, OWD2025Estimated, worldometers}. 
These cumulative records are transformed into compartmental trajectories consistent with the SEIRD framework. 
Specifically, the infected compartment ($I$) is reconstructed from active case estimates, the deceased compartment ($D$) is reconstructured from reported mortality data, and the recovered compartment ($R$) is reconstructed by aggregating recovered and deceased cases when recovery information is available.
Since the exposed compartment ($E$) is not directly observable in the data, it is treated as a latent state variable to be inferred by the proposed model.

\begin{figure}[H]
\centering
\begin{subfigure}[b]{\textwidth}
\centering
\includegraphics[width=1\textwidth]{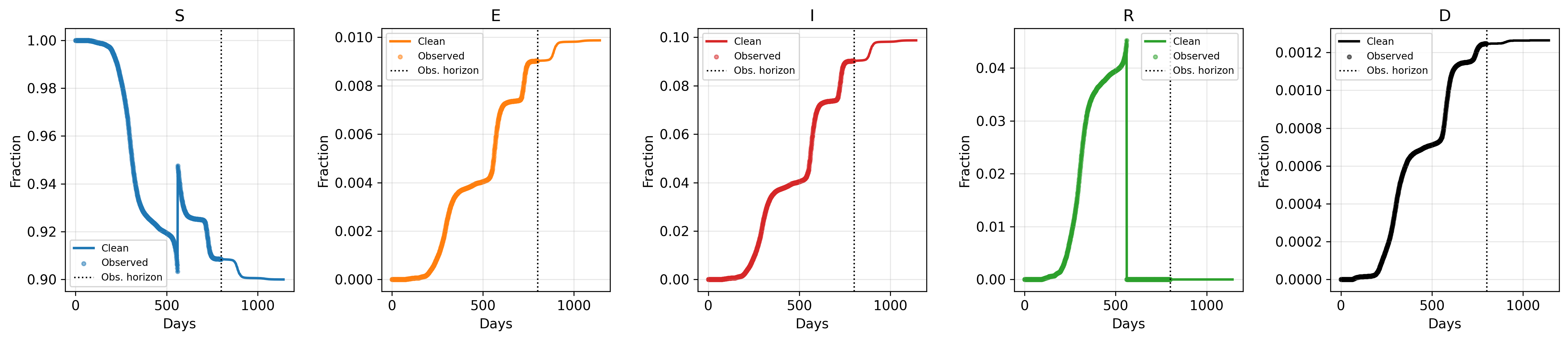}
\caption{Germany}\label{F9a}
\end{subfigure}
\hfill
\begin{subfigure}[b]{\textwidth}
\centering
\includegraphics[width=1\textwidth]{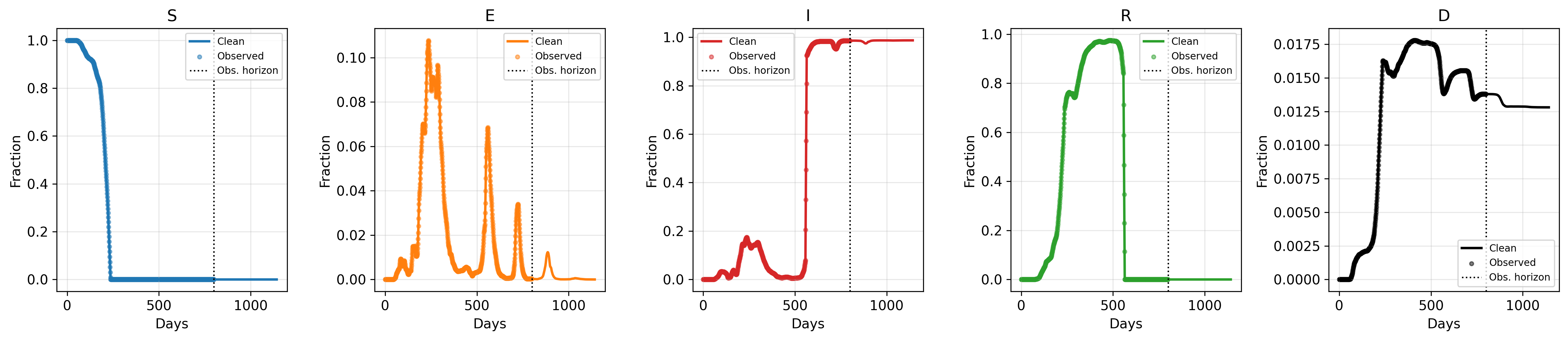}
\caption{Morocco}\label{F9b}
\end{subfigure}
\hfill
\begin{subfigure}[b]{\textwidth}
\centering
\includegraphics[width=\textwidth]{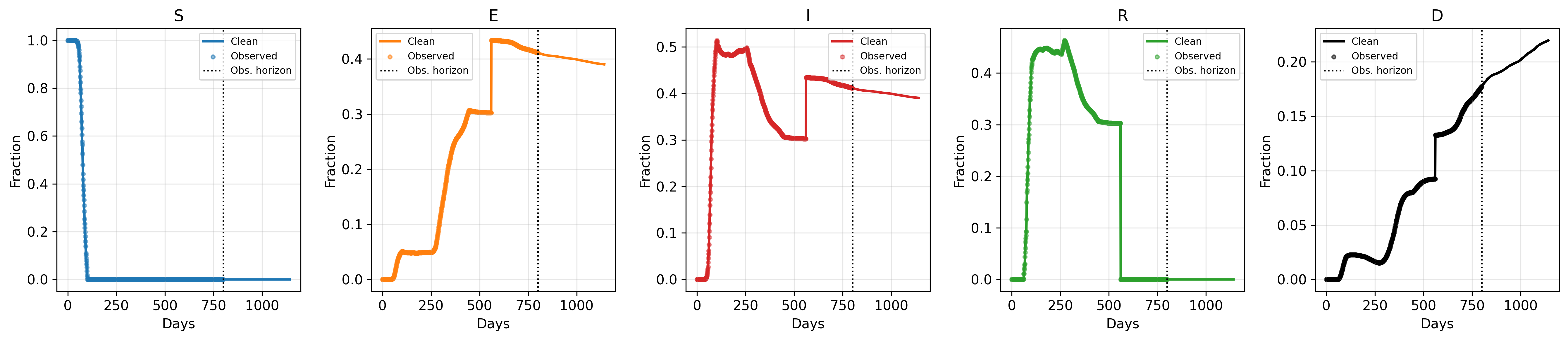}
\caption{Sweden}\label{F9c}
\end{subfigure}
\caption{Real-world COVID-19 epidemic trajectories for Germany, Morocco, and Sweden.}\label{F9}
\end{figure}

Figure~\ref{F9} presents the real-world trajectories of the spread of COVID-19 in Germany, Morocco, and Sweden, which were used in the numerical experiments. 
The compartments representing the susceptible, exposed, infected, recovered, and deceased illustrate the different epidemic patterns in these countries and provide a benchmark for evaluating the reconstruction, parameter estimation, and forecasting capabilities of the proposed framework.

\begin{figure}[H]
\centering
\begin{subfigure}[b]{\textwidth}
\centering
\includegraphics[width=1\textwidth]{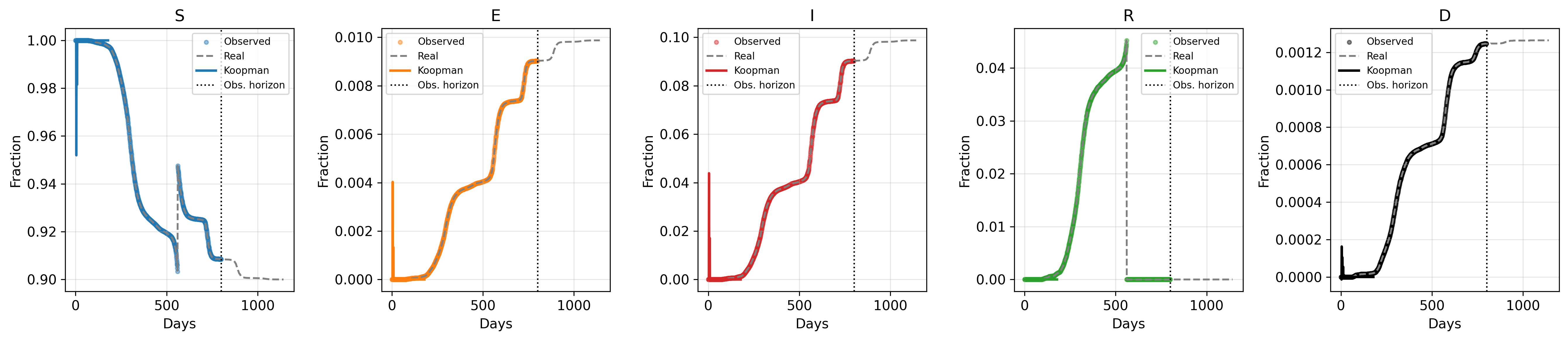}
\caption{Germany}\label{F10a}
\end{subfigure}
\hfill
\begin{subfigure}[b]{\textwidth}
\centering
\includegraphics[width=1\textwidth]{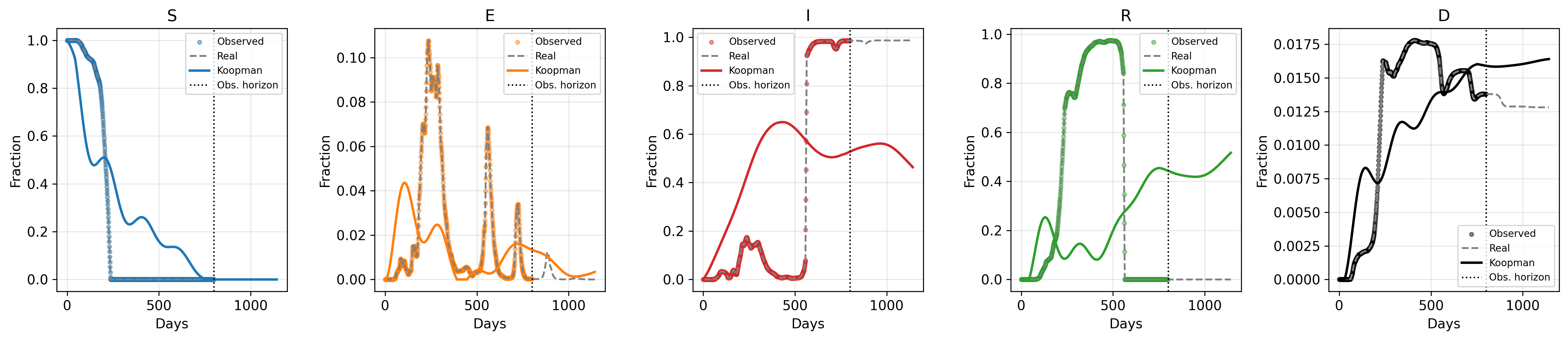}
\caption{Morocco}\label{F10b}
\end{subfigure}
\hfill
\begin{subfigure}[b]{\textwidth}
\centering
\includegraphics[width=\textwidth]{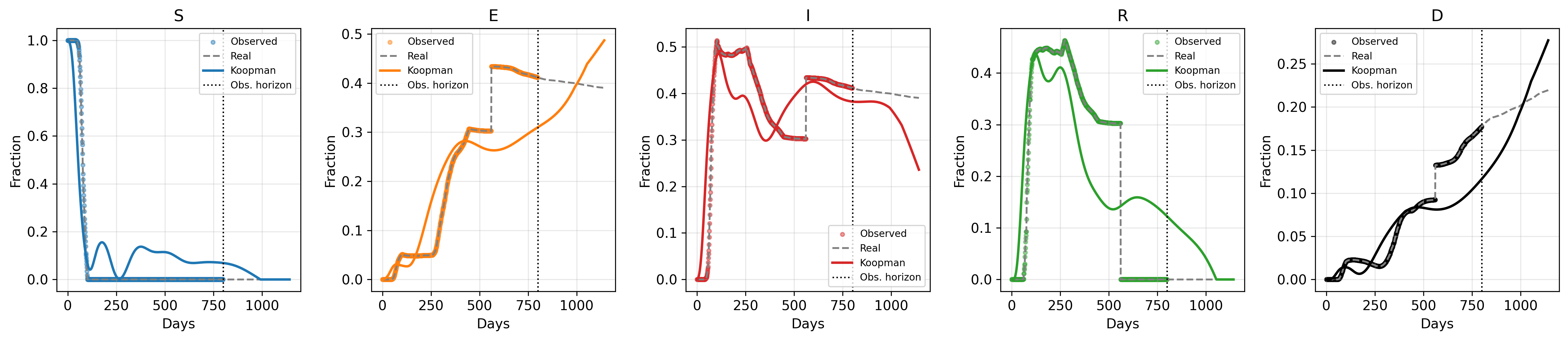}
\caption{Sweden}\label{F10c}
\end{subfigure}
\caption{Comparison between the observed epidemiological trajectories and the K--EDMD reconstructions for the Germany, Morocco, and Sweden datasets.}\label{F10}
\end{figure}

Figure~\ref{F10} presents the reconstruction and forecasting results obtained using the K--EDMD approach for the Germany, Morocco, and Sweden datasets. 
Although the Koopman-based representation captures the overall qualitative trends of the epidemic evolution, noticeable discrepancies remain in several compartments, especially during the forecasting interval. 
These results suggest that the finite-dimensional Koopman approximation alone is insufficient for accurately representing the complex nonlinear interactions that govern the epidemic dynamics. 
The corresponding quantitative errors reported in Table~\ref{Tab5} confirm the limited predictive performance of the standalone K--EDMD approach.

\begin{figure}[H]
\centering
\begin{subfigure}[b]{\textwidth}
\centering
\includegraphics[width=1\textwidth]{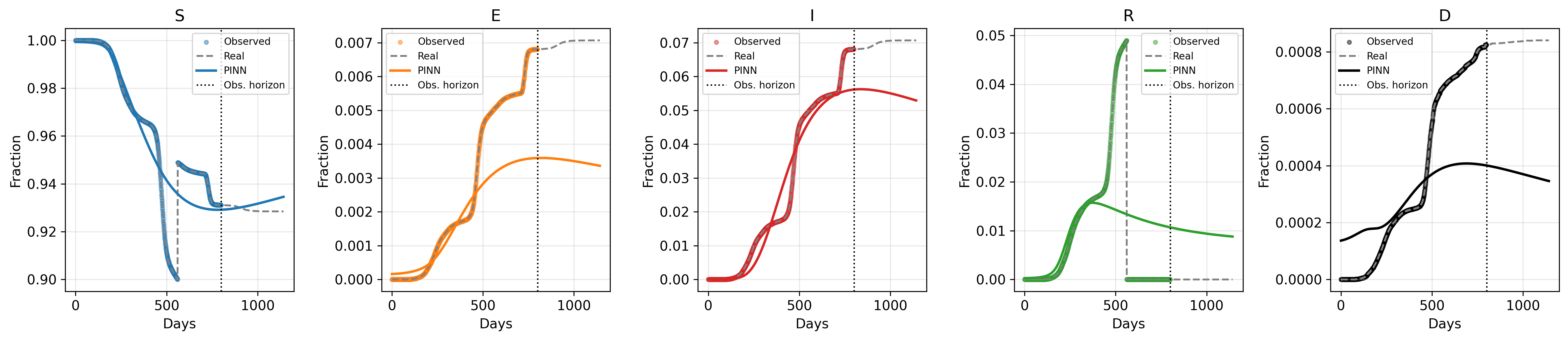}
\caption{Germany}\label{F11a}
\end{subfigure}
\hfill
\begin{subfigure}[b]{\textwidth}
\centering
\includegraphics[width=1\textwidth]{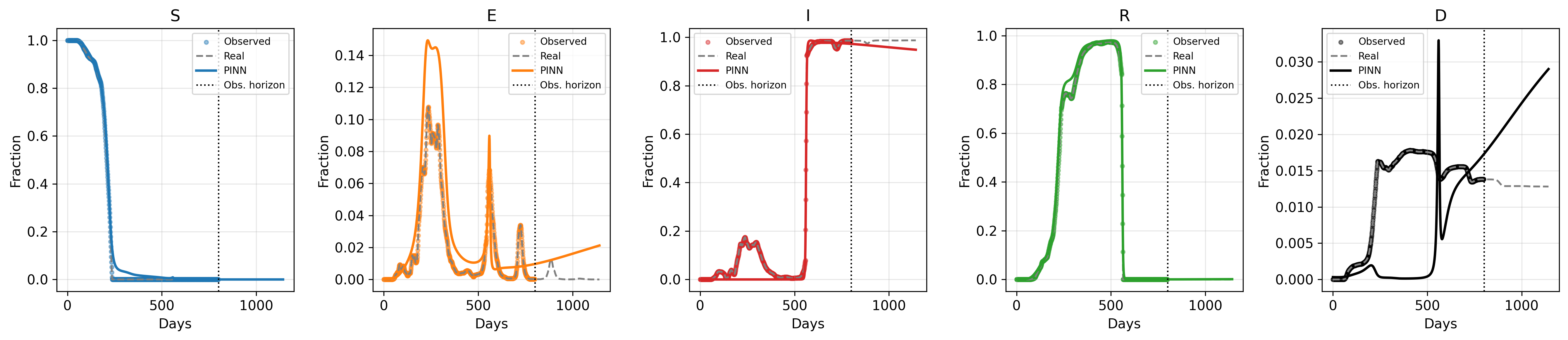}
\caption{Morocco}\label{F11b}
\end{subfigure}
\hfill
\begin{subfigure}[b]{\textwidth}
\centering
\includegraphics[width=\textwidth]{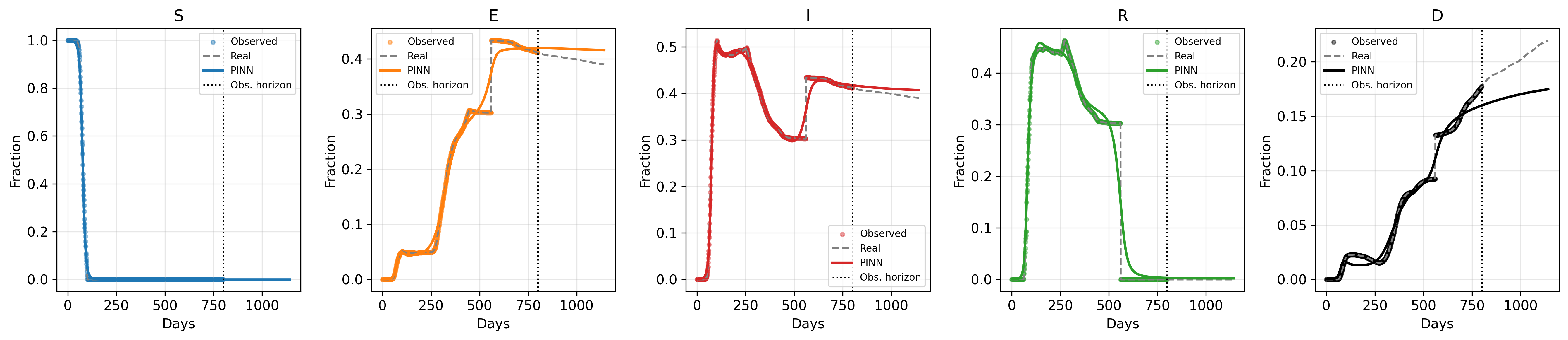}
\caption{Sweden}\label{F11c}
\end{subfigure}
\caption{Comparison between the observed epidemiological trajectories and the PINN reconstructions for the Germany, Morocco, and Sweden datasets.}\label{F11}
\end{figure}

Figure~\ref{F11} illustrates the reconstruction and forecasting performance of the PINN framework for the three real-world datasets. 
By incorporating the governing epidemic equations into the learning process, the PINN framework significantly improves the agreement between the predicted and observed trajectories compared with the K--EDMD approach. 
The reconstructed solutions remain stable throughout both the training and forecasting intervals,  accurately capturing the main epidemic trends. 
However, small deviations are still visible in some compartments, particularly during long-term forecasting, reflecting the challenges of learning highly nonlinear epidemic dynamics from limited observations.
The corresponding quantitative errors are reported in Table~\ref{Tab5}.

\begin{figure}[H]
\centering
\begin{subfigure}[b]{\textwidth}
\centering
\includegraphics[width=1\textwidth]{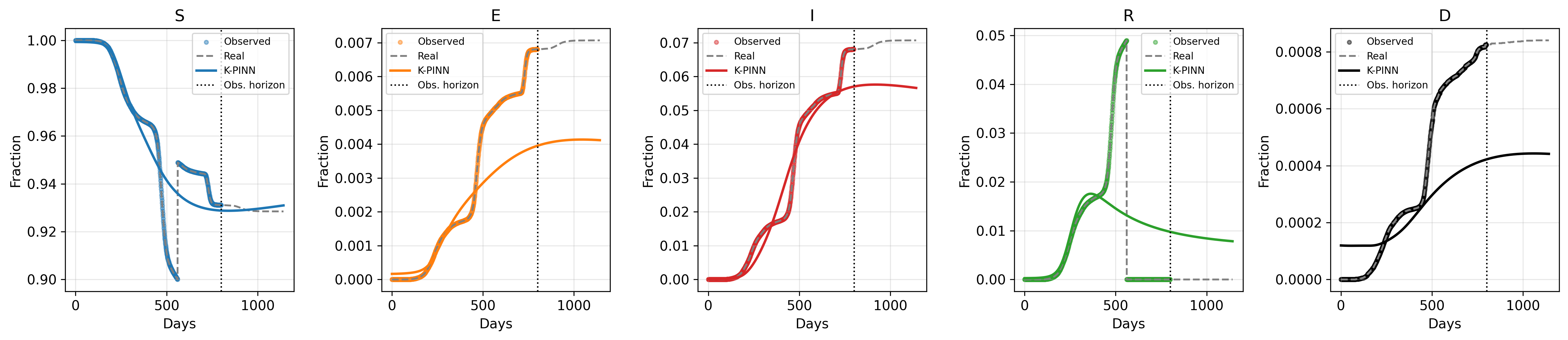}
\caption{Germany}\label{F12a}
\end{subfigure}
\hfill
\begin{subfigure}[b]{\textwidth}
\centering
\includegraphics[width=1\textwidth]{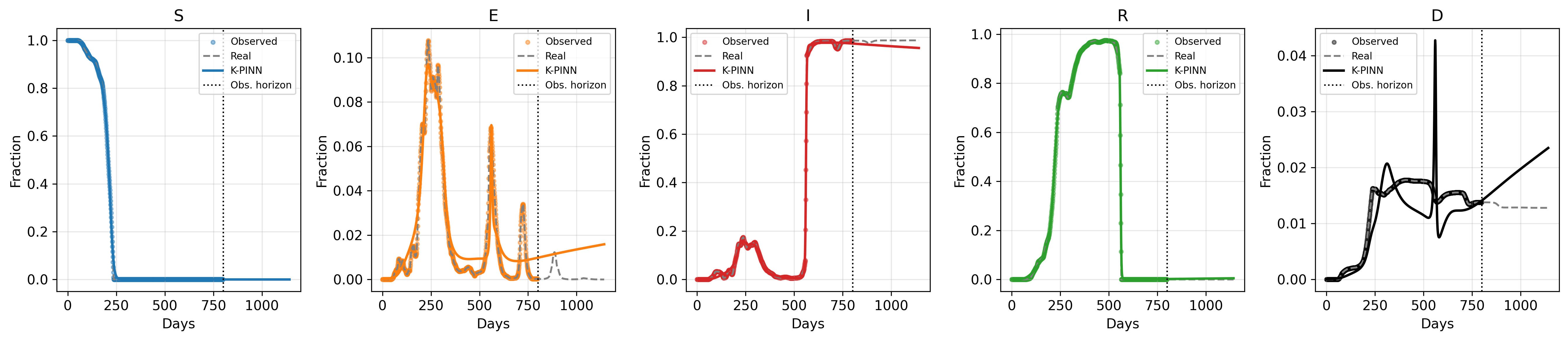}
\caption{Morocco}\label{F12b}
\end{subfigure}
\hfill
\begin{subfigure}[b]{\textwidth}
\centering
\includegraphics[width=\textwidth]{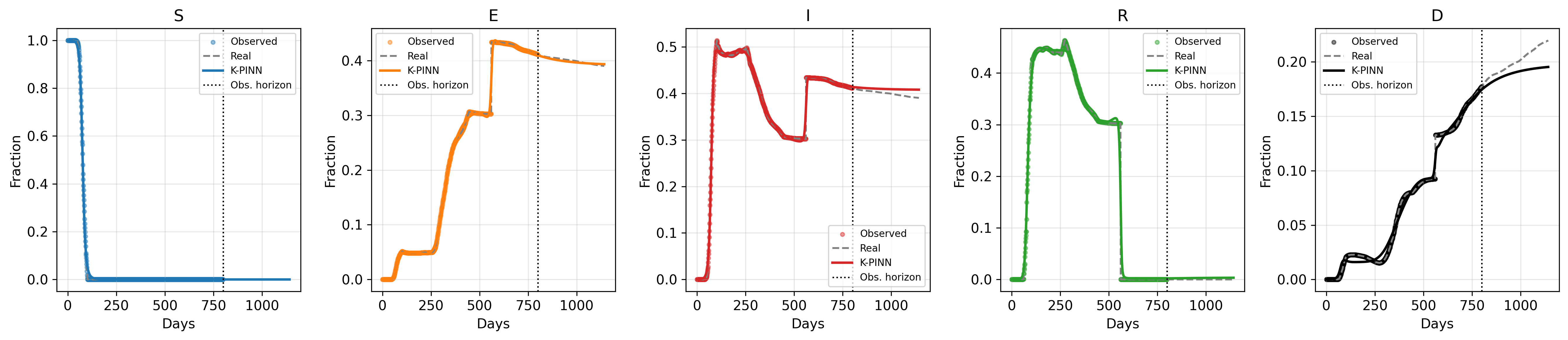}
\caption{Sweden}\label{F12c}
\end{subfigure}
\caption{Comparison between the observed epidemiological trajectories and the proposed K--PINN reconstructions for the Germany, Morocco, and Sweden datasets.}\label{F12}
\end{figure}

Figure~\ref{F12} presents the reconstruction and forecasting results obtained using the proposed K--PINN framework. 
The predicted trajectories show excellent alignment with the observed epidemic data across all compartments and datasets. 
Compared with both K--EDMD and PINN, the proposed approach provides more accurate reconstructions and improved long-term forecasting performance. 
These results demonstrate that the combining Koopman latent linearization and physics-informed learning effectively captures the underlying epidemic dynamics while preserving the model's mechanistic structure.

The quantitative results reported in Tables~\ref{Tab5} and \ref{Tab6} further confirm the advantages of the proposed K--PINN framework. 
For all three datasets, the K--PINN framework achieves the lowest global reconstruction errors, outperforming both the K--EDMD and the classical PINN approach. 
This improvement is particularly noticeable for the Morocco and Sweden datasets, where the nonlinear epidemic dynamics are more challenging to capture. 
These results demonstrate that integrating Koopman-based latent representations with physics-informed learning results in more precise state reconstruction, enhanced forecasting capabilities, and reliable epidemiological parameter estimation.
\begin{table}[H]
\centering
\caption{Comparison of reconstruction and forecasting errors obtained using K--EDMD, PINN, and the proposed K--PINN framework for the Germany, Morocco, and Sweden datasets.}\label{Tab5}
\resizebox{\textwidth}{!}{
\begin{tabular}{ccccccccc}
\hline
\textbf{Dataset} & \textbf{Method} & $E_{L^2}$ & $E_{L^\infty}$ & RMSE$(S)$ & RMSE$(E)$ & RMSE$(I)$ & RMSE$(R)$ & RMSE$(D)$\\
\hline\hline
\multirow{3}{*}{Germany}
& K--EDMD & 0.89466 & 0.99964 & 0.89319 & 0.00528 & 0.04902 & 0.01421 & 0.00060 \\
& PINN   & 0.01830 & 0.03597 & 0.01067 & 0.00225 & 0.00927 & 0.01140 & 0.00031 \\
& K--PINN & 0.01731 & 0.03584 & 0.01055 & 0.00194 & 0.00800 & 0.01098 & 0.00029 \\
\hline
\multirow{3}{*}{Morocco}
& K--EDMD & 0.70428 & 0.88771 & 0.19012 & 0.02544 & 0.44309 & 0.51271 & 0.00387 \\
& PINN   & 0.06419 & 0.17455 & 0.01948 & 0.02161 & 0.05218 & 0.02084 & 0.01083 \\
& K--PINN & 0.02278 & 0.11854 & 0.00727 & 0.00986 & 0.01589 & 0.00949 & 0.00513 \\
\hline
\multirow{3}{*}{Sweden}
& K--EDMD & 0.19763 & 0.53042 & 0.11379 & 0.08303 & 0.07483 & 0.11130 & 0.03505 \\
& PINN   & 0.03824 & 0.15156 & 0.00592 & 0.01684 & 0.01303 & 0.02507 & 0.01858 \\
& K--PINN & 0.01812 & 0.14256 & 0.00591 & 0.00528 & 0.00865 & 0.01111 & 0.00822 \\
\hline
\end{tabular}}
\end{table}
\begin{table}[H]
\centering
\setlength{\tabcolsep}{0.5cm}
\caption{Estimated epidemiological parameters obtained using PINN and the proposed K--PINN framework for the real-world datasets.}\label{Tab6}
\begin{tabular}{cccccc}
\hline
\textbf{Dataset} & \textbf{Method} & $\beta$ & $\sigma$ & $\gamma$ & $\mu$\\
\hline\hline
\multirow{2}{*}{Germany}
& PINN   & 0.2207 & 0.2595 & 0.0576 & 0.0039 \\
& K--PINN & 0.2254 & 0.2596 & 0.0589 & 0.0041 \\
\hline
\multirow{2}{*}{Morocco}
& PINN   & 0.3096 & 0.1454 & 0.0547 & 0.0019 \\
& K--PINN & 0.2593 & 0.1744 & 0.0547 & 0.0019 \\
\hline
\multirow{2}{*}{Sweden}
& PINN   & 0.2583 & 0.1484 & 0.0620 & 0.0075 \\
& K--PINN & 0.2257 & 0.1200 & 0.0544 & 0.0078 \\
\hline
\end{tabular}
\end{table}

For clarity of presentation, Figures~\ref{FB1}--\ref{FB3} display the collective reconstruction and forecasting results obtained with the proposed K--PINN framework. 
Detailed compartment-wise comparisons for all epidemic variables and all competing approaches are provided in Appendix~\ref{App2}.

\section{Conclusion and Future Work}\label{S6}
In this paper, we proposed a K--PINN framework for nonlinear epidemic systems. 
This methodology combines the strengths of Koopman analysis, an operator-theoretic approach, with the capabilities of PINNs, which enforce equations to provide a robust, interpretable, and data-efficient framework for epidemic reconstruction, parameter inference, and forecasting. 
The approach elevates the epidemic dynamics to a latent observable space, where evolution is approximately governed by a finite-dimensional Koopman operator. 
Through a physics-informed loss function, the framework simultaneously enforces the governing epidemic equations and integrates data-driven learning with mechanistic epidemiological modeling.

We developed and analyzed the framework using an SEIRSD epidemic model that incorporates disease-induced mortality and temporary immunity. 
We presented theoretical considerations on the model's well-posedness and invariance properties, providing a mathematically consistent foundation for the subsequent Koopman-based learning framework. 
Next, we introduced a unified K--PINN architecture consisting of a neural lifting operator, latent Koopman dynamics, a reconstruction network, and physics-informed regularization terms that enforce the governing epidemic equations and epidemiological constraints.

A series of numerical experiments involving both synthetic and real-world epidemiological datasets demonstrated the effectiveness of the proposed methodology. 
The synthetic experiments confirmed the framework's ability to accurately reconstruct epidemic trajectories and identify unknown epidemiological parameters.
Additional validation using COVID-19 datasets from Germany, Morocco, and Sweden further demonstrated the applicability of the proposed approach to real epidemic observations. 
In all scenarios considered, the K--PINN framework achieved improved reconstruction accuracy, enhanced parameter identifiability, and superior forecasting performance compared to the standalone K--EDMD approach and the classical PINN framework. 
These results suggest that combining Koopman latent linearization with physics-informed learning is a powerful way to capture complex nonlinear epidemic dynamics while preserving the underlying epidemiological structure.

The proposed framework offers several important advantages. 
First, the latent Koopman representation provides a structured and interpretable description of nonlinear dynamics, thereby improving long-term forecasting capabilities. 
Second, physics-informed constraints ensure consistency with the governing epidemic model and reduce the risk of physically unrealistic predictions. 
Third, the unified inverse-learning formulation enables simultaneous state reconstruction and parameter estimation from limited observations. This makes the approach particularly attractive for real-world epidemic monitoring and decision support.


The present work yields several promising directions for future research. 
From an epidemiological perspective, for example, the proposed K--PINN framework can be extended to more sophisticated disease transmission models. 
These models can include age-structured, multi-group, multi-strain, and metapopulation systems.
 This would enable the analysis of heterogeneous populations, spatial mobility, and complex contact networks. 
Another important avenue is incorporating stochastic effects to account for environmental variability, demographic randomness, and measurement uncertainties. 
Additionally, the methodology can be generalized to fractional-order epidemic models, enabling memory and hereditary effects to be explicitly represented within the learning framework. 

The proposed K--PINN methodology has applications beyond epidemiology and can be used with a broad range of nonlinear dynamical systems. 
Potential applications include ecological population dynamics, predator-prey systems, tumor growth models, biological networks, reaction-diffusion processes in chemistry, fluid dynamics, climate and environmental modeling, and engineering systems governed by ordinary, fractional, stochastic, or partial differential equations. 
These extensions demonstrate the versatility of the proposed framework as a general machine-learning methodology for complex nonlinear systems, combining latent linear representations with physics-informed learning.

\begin{appendices}
\section{Additional and Supplementary Synthetic Data Experiments}\label{App1}
\begin{figure}[H]
\centering
\includegraphics[width=1\textwidth]{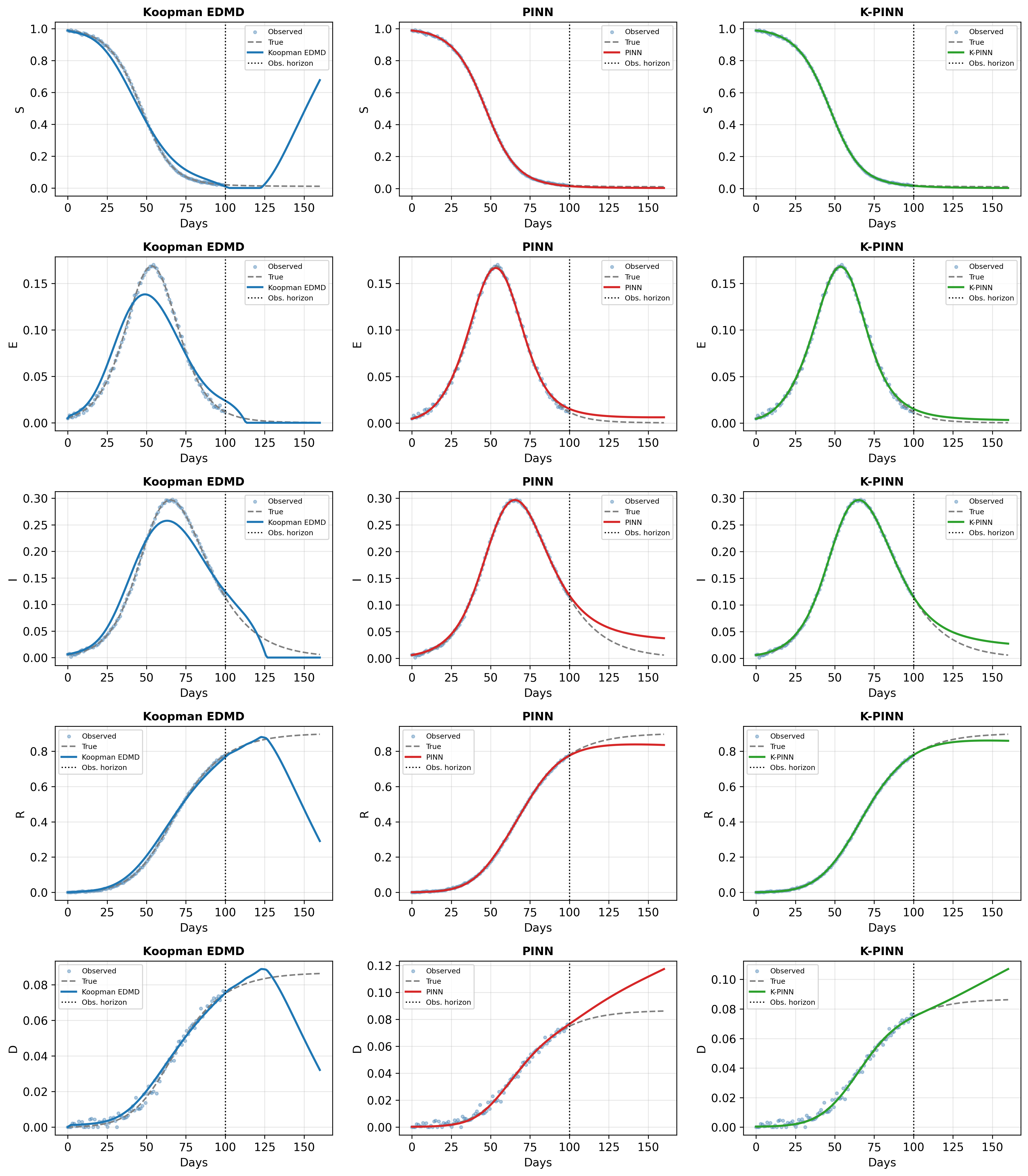}
\caption{Compartment-wise comparison of K--EDMD, PINN, and K--PINN predictions against the synthetic trajectories.}\label{FA1}
\end{figure}

This appendix provides a detailed, compartment-by-compartment comparison of the reconstruction and forecasting results obtained using K--EDMD, PINN, and the proposed K--PINN framework. 
Figure~\ref{FA1} compares the reference synthetic trajectory with the corresponding predictions generated by the three approaches for each epidemiological compartment over the training and forecasting intervals. 
These supplementary results provide insight into the strengths and limitations of each methodology, further highlighting the improved accuracy and long-term predictive capability of the K--PINN framework.

\begin{figure}[H]
\centering
\includegraphics[width=1\textwidth]{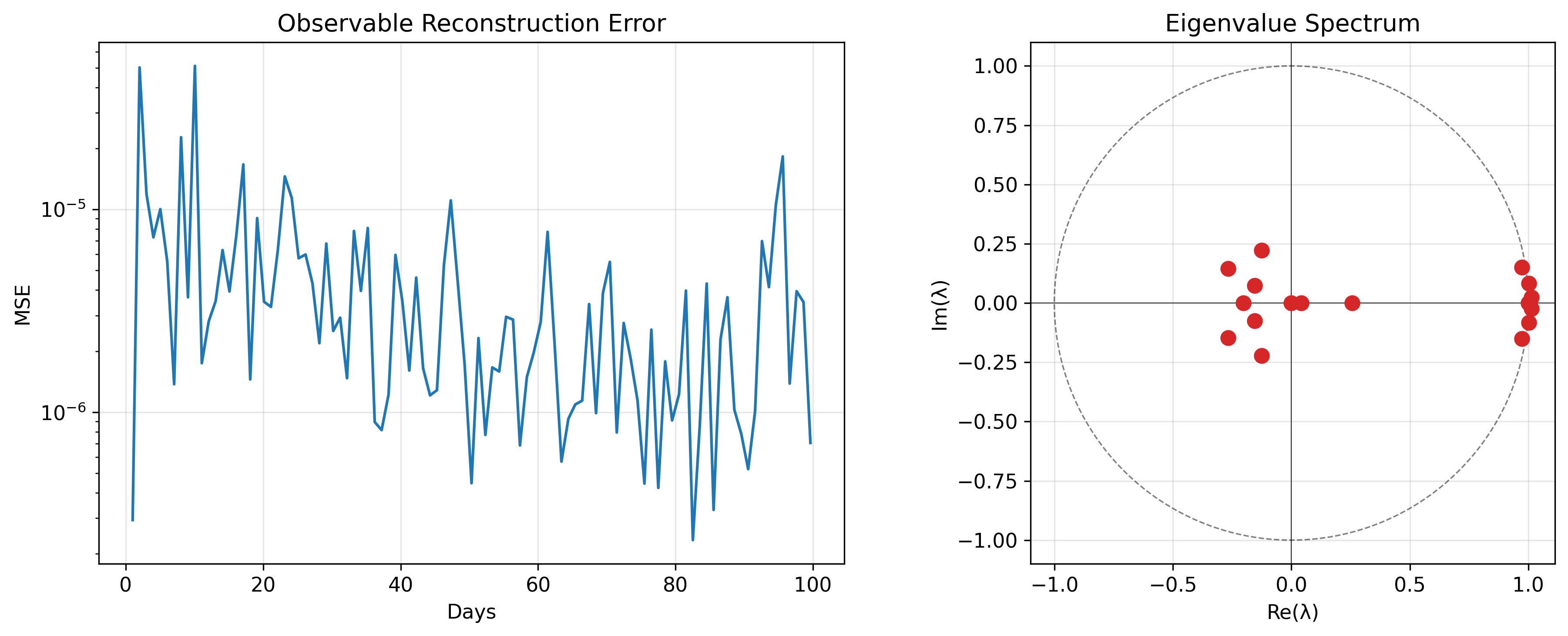}
\caption{K--EDMD diagnostic analysis and latent-space prediction errors for the synthetic dataset.}\label{FA2}
\end{figure}

Figure~\ref{FA2}  reports the diagnostic quantities associated with the Koopman approximation, including the latent-space prediction errors and the reconstruction diagnostics. 

\begin{figure}[H]
\centering
\includegraphics[width=1\textwidth]{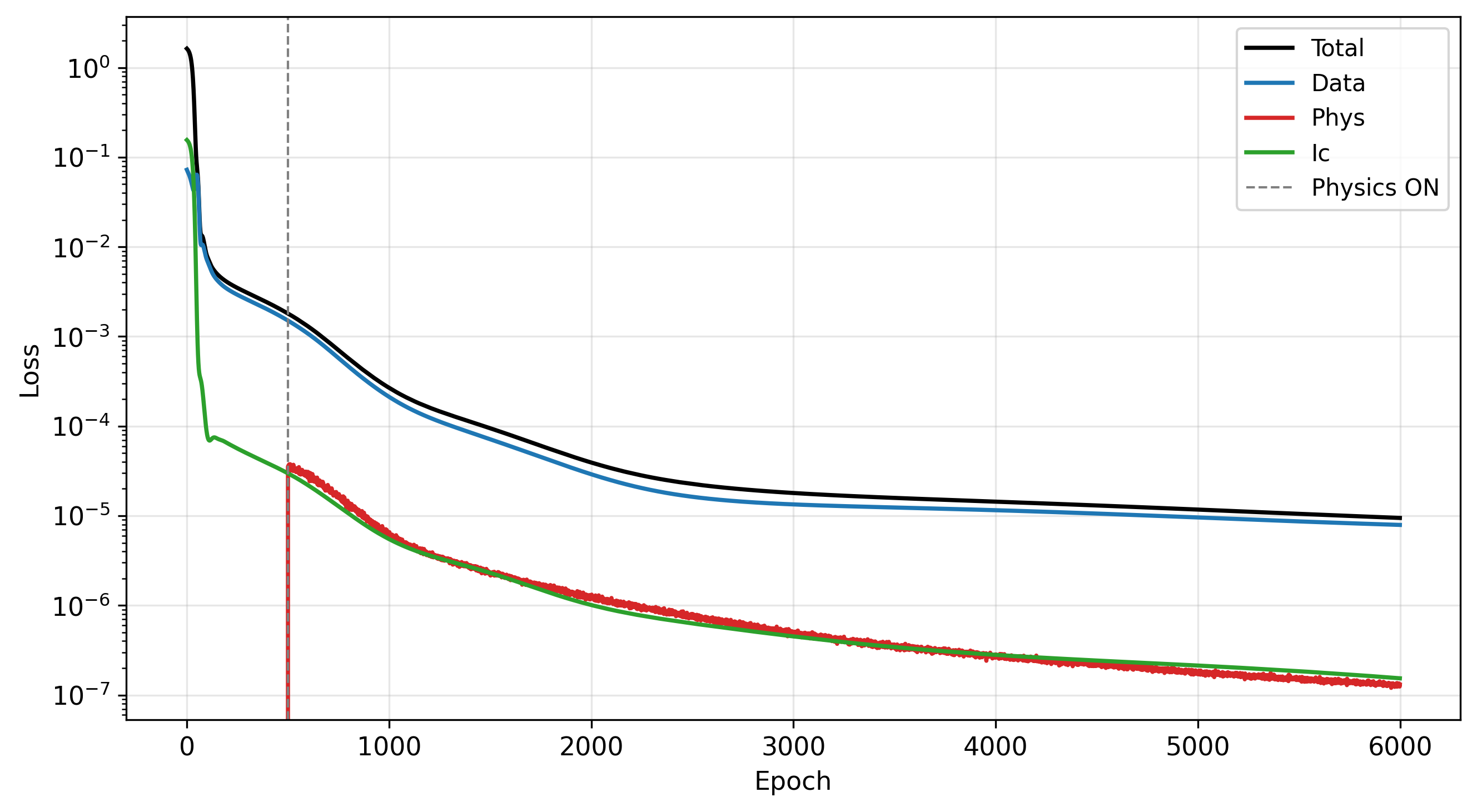}
\caption{Evolution of the composite loss function during PINN training for the synthetic dataset.}\label{FA3}
\end{figure}

\begin{figure}[H]
\centering
\includegraphics[width=1\textwidth]{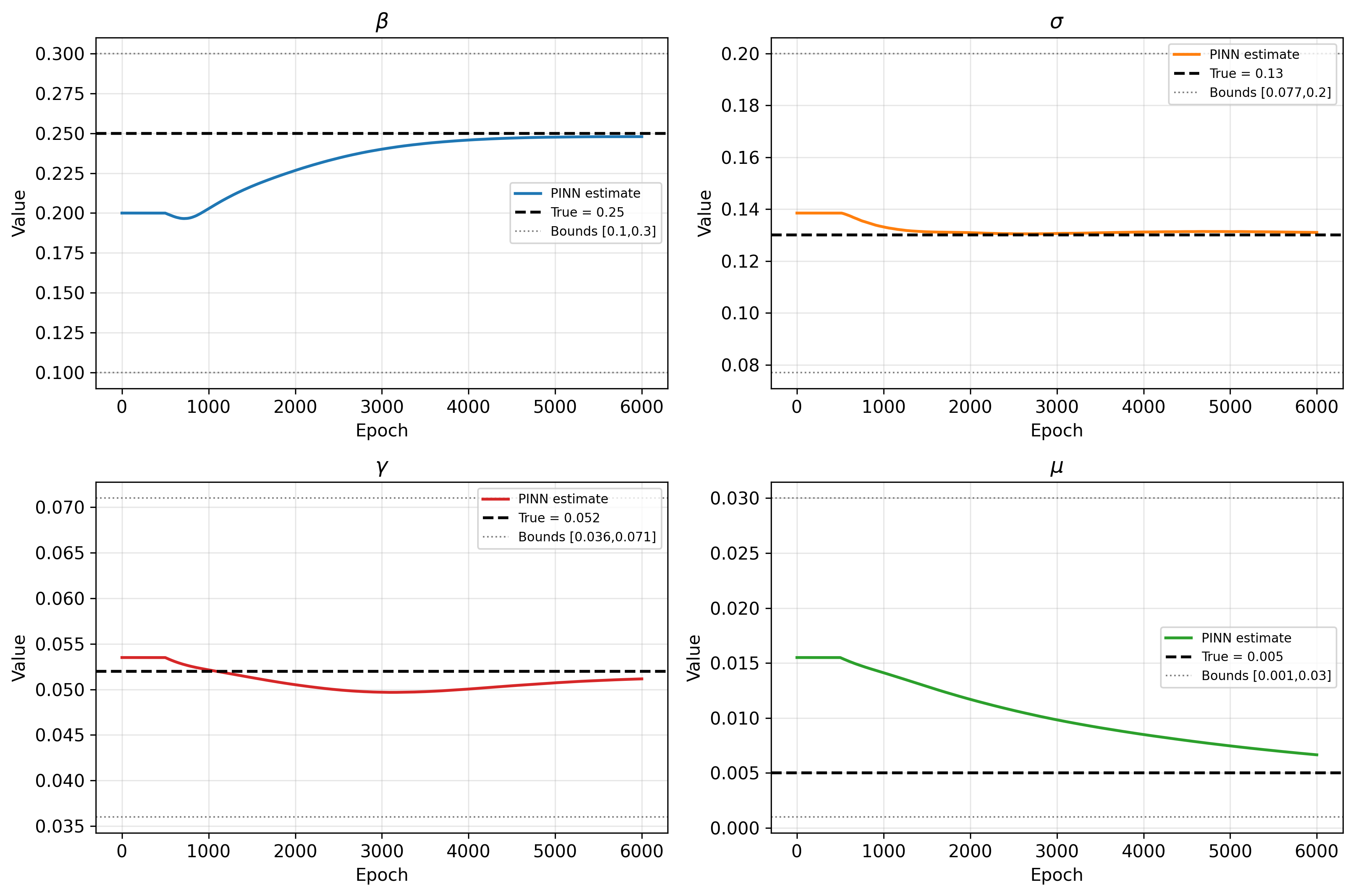}
\caption{Convergence of epidemiological parameter estimates obtained by PINN for the synthetic dataset.}\label{FA4}
\end{figure}

Figure~\ref{FA3} shows the convergence history of the PINN optimization process, and Figure~\ref{FA4}  shows the evolution of the estimated epidemiological parameters during training. 

\section{Additional and Supplementary Real-World Data Experiments}\label{App2}
\begin{figure}[H]
\centering
\includegraphics[width=1\textwidth]{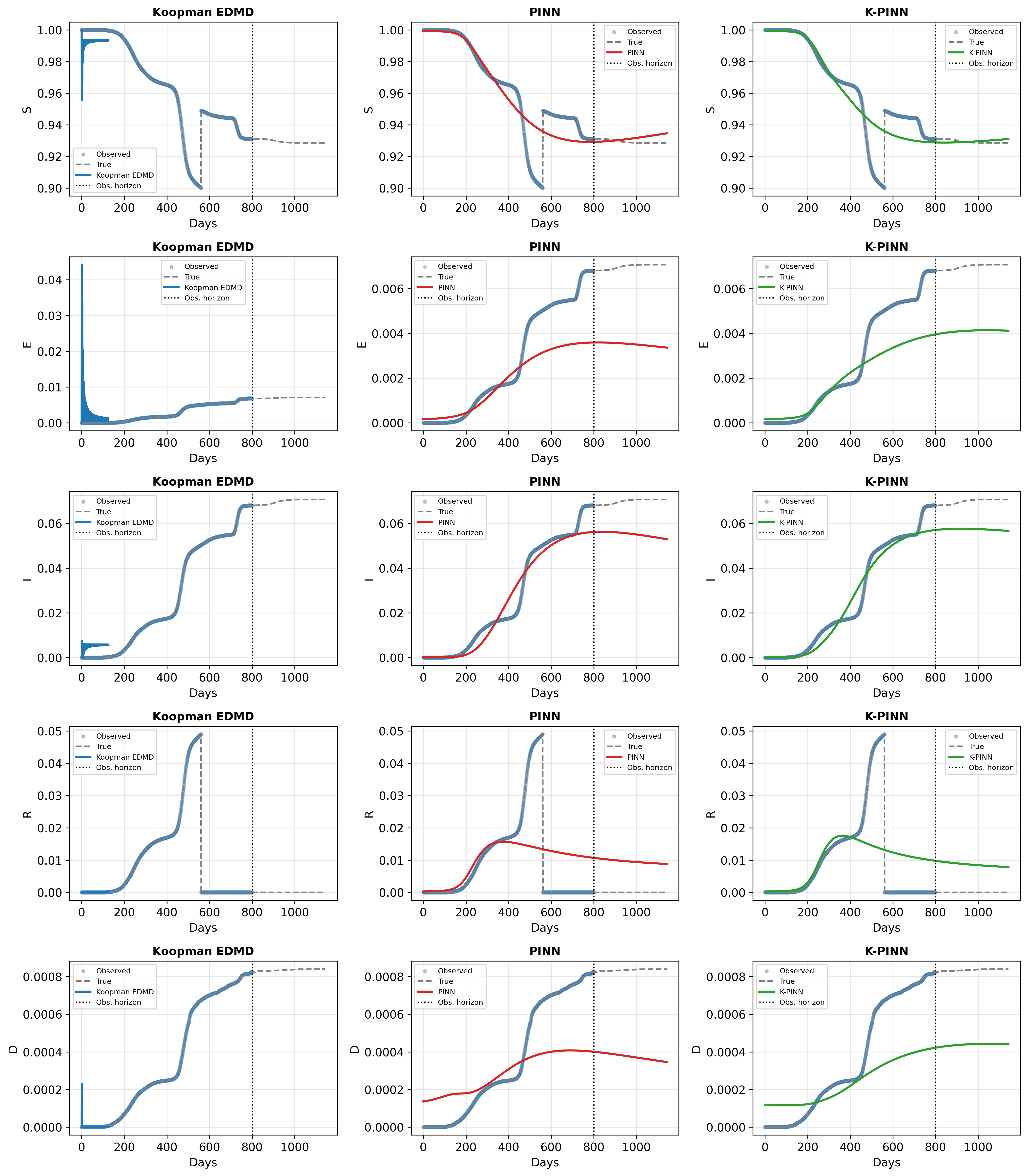}
\caption{Compartment-wise comparison of K--EDMD, PINN, and K--PINN predictions for the COVID-19 Germany dataset.}\label{FB1}
\end{figure}

\begin{figure}[H]
\centering
\includegraphics[width=1\textwidth]{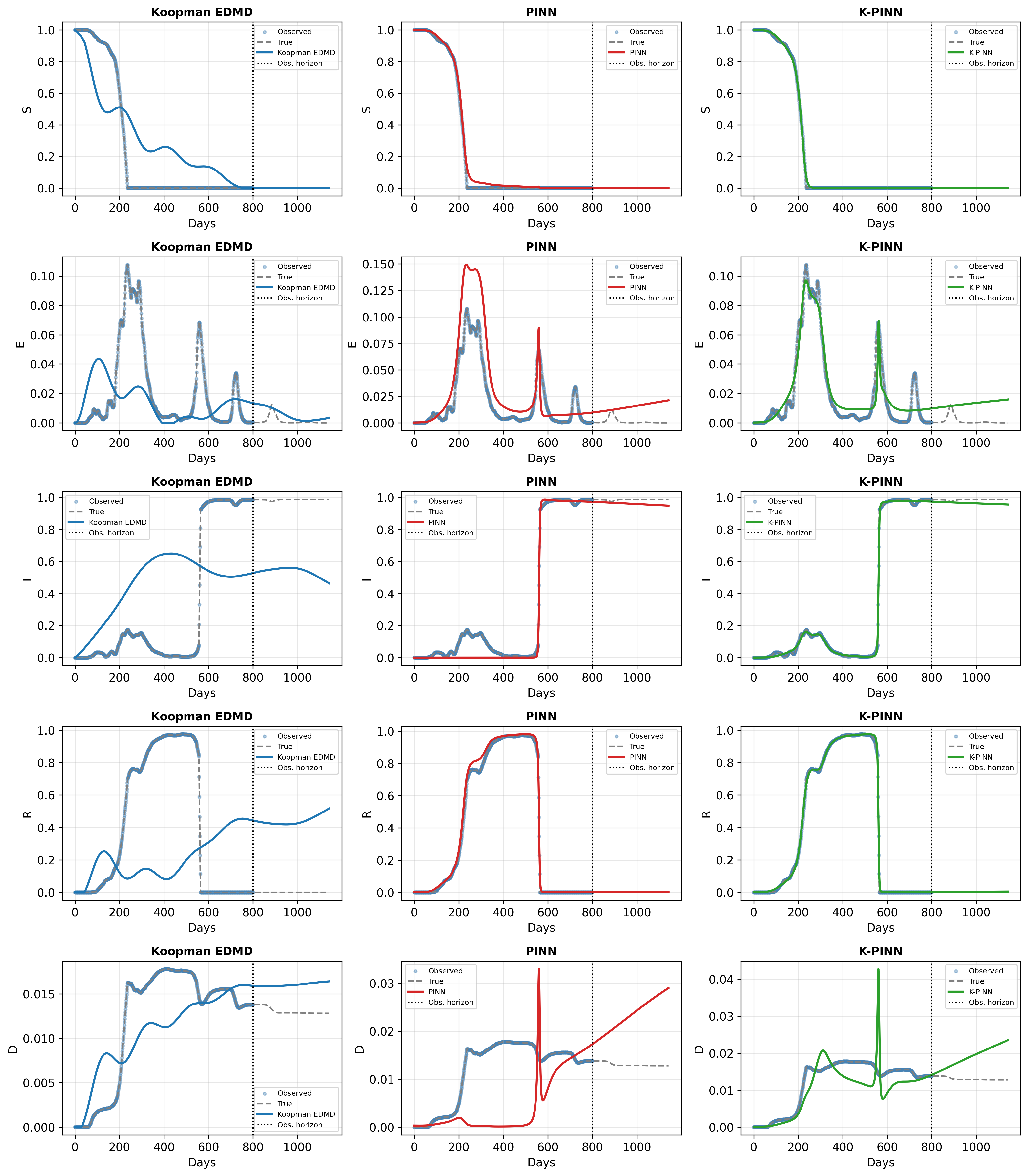}
\caption{Compartment-wise comparison of K--EDMD, PINN, and K--PINN predictions for the COVID-19 Morocco dataset.}\label{FB2}
\end{figure}

\begin{figure}[H]
\centering
\includegraphics[width=1\textwidth]{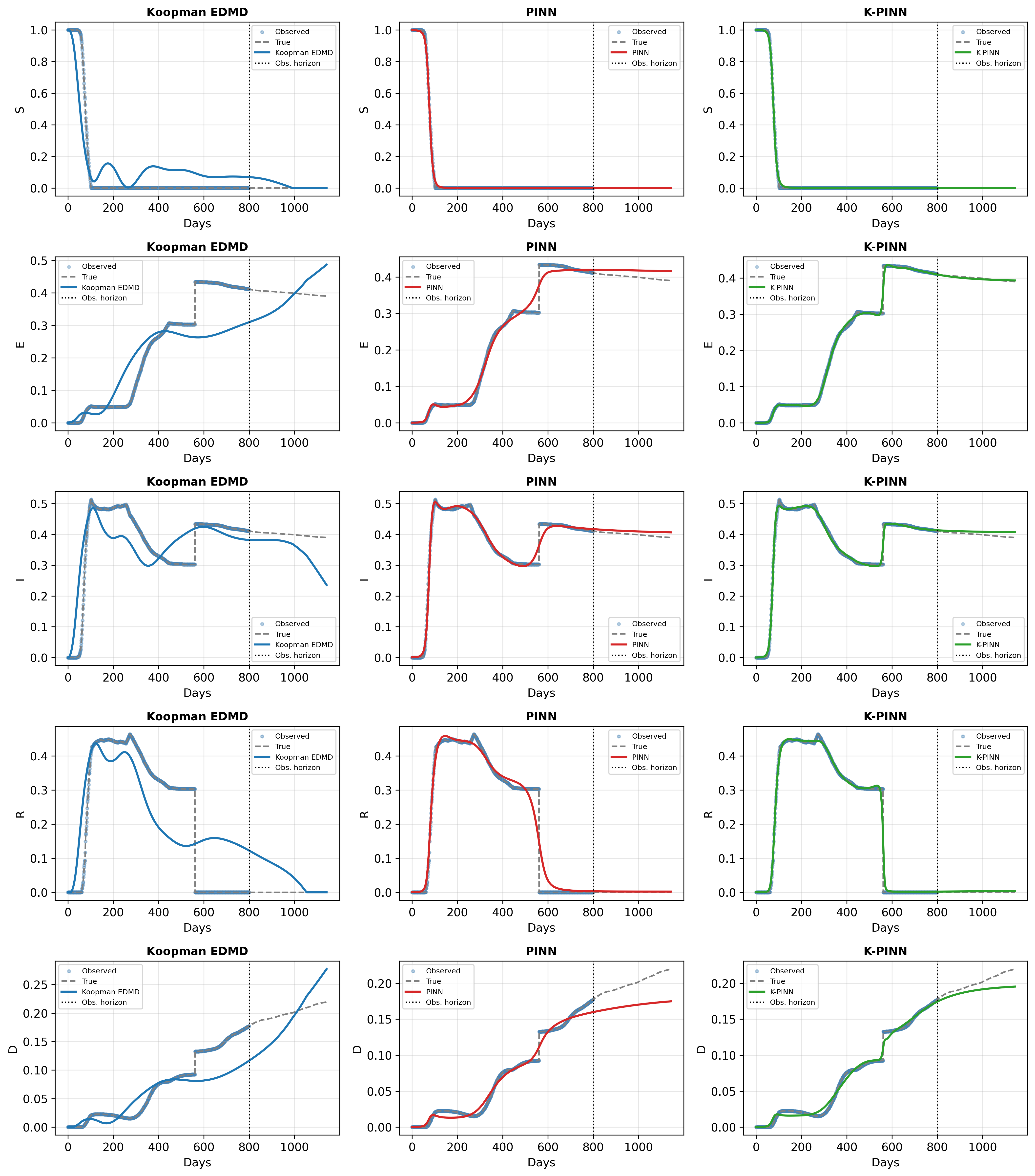}
\caption{Compartment-wise comparison of K--EDMD, PINN, and K--PINN predictions for the COVID-19 Sweden dataset.}\label{FB3}
\end{figure}

This appendix provides a detailed, compartment-by-compartment comparison of the reconstruction and forecasting results obtained using K--EDMD, PINN, and the proposed K--PINN framework for the Germany, Morocco, and Sweden datasets.
Figures~\ref{FB1}--\ref{FB3} compare the observed epidemic trajectories with the corresponding predictions generated by the three approaches over the training and forecasting intervals for each epidemiological compartment.
These supplementary results provide insight into the strengths and limitations of each methodology, further highlighting the improved accuracy, robustness, and long-term predictive capability of the proposed K--PINN framework.

\begin{figure}[H]
\centering
\begin{subfigure}[b]{.5\textwidth}
\centering
\includegraphics[width=1\textwidth]{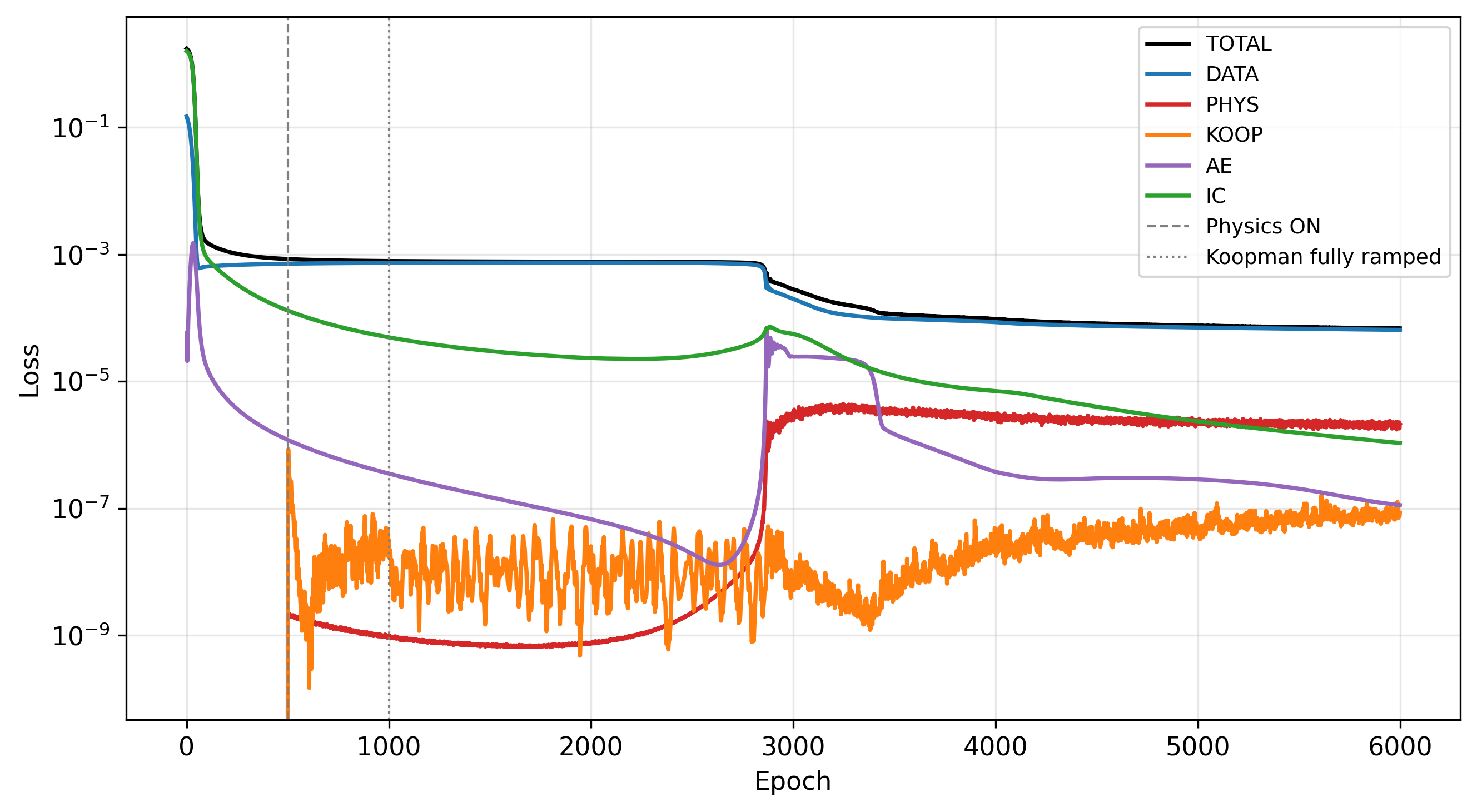}
\caption{Germany}\label{FB4a}
\end{subfigure}%
~
\begin{subfigure}[b]{.5\textwidth}
\centering
\includegraphics[width=1\textwidth]{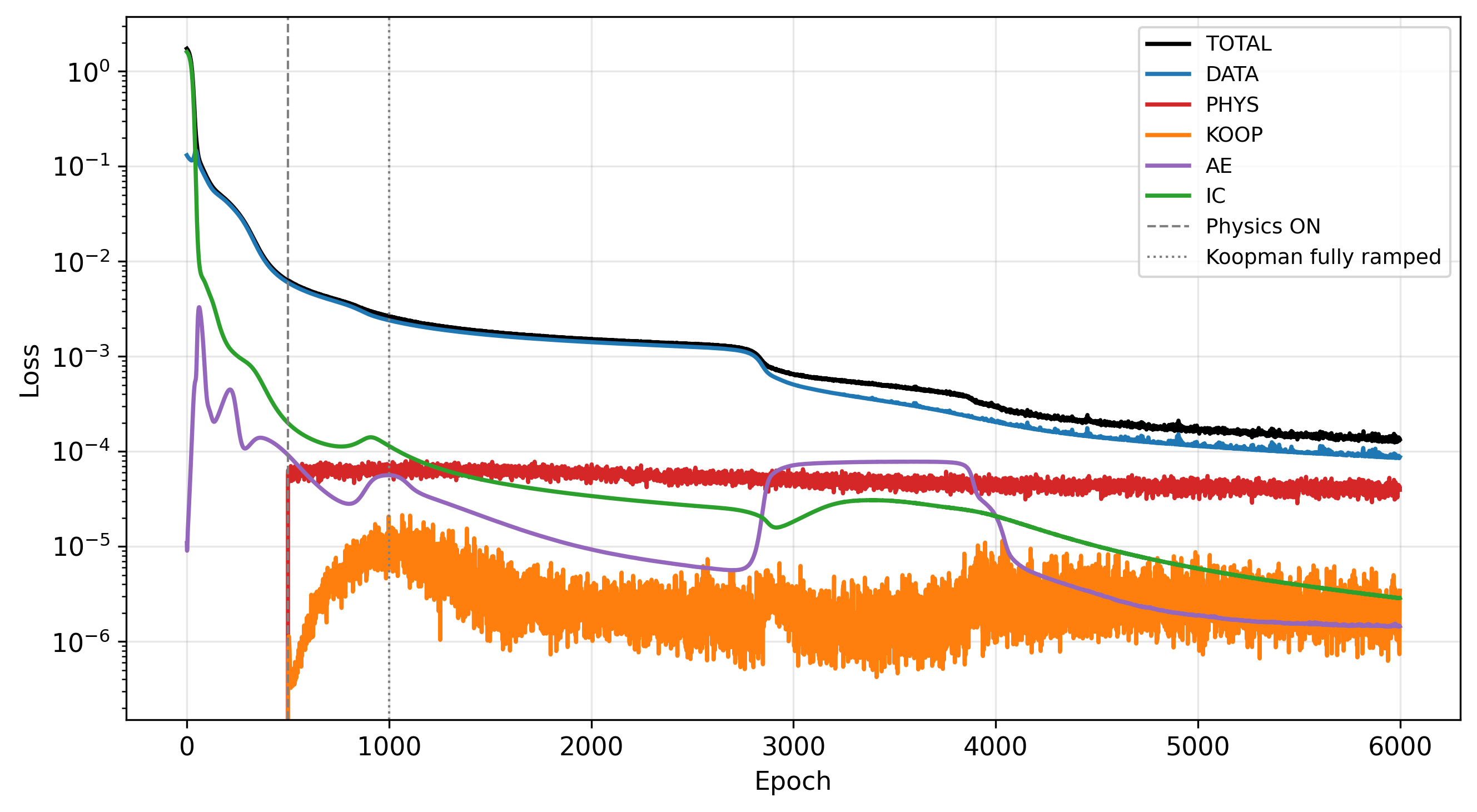}
\caption{Morocco}\label{FB4b}
\end{subfigure}
~
\begin{subfigure}[b]{.5\textwidth}
\centering
\includegraphics[width=\textwidth]{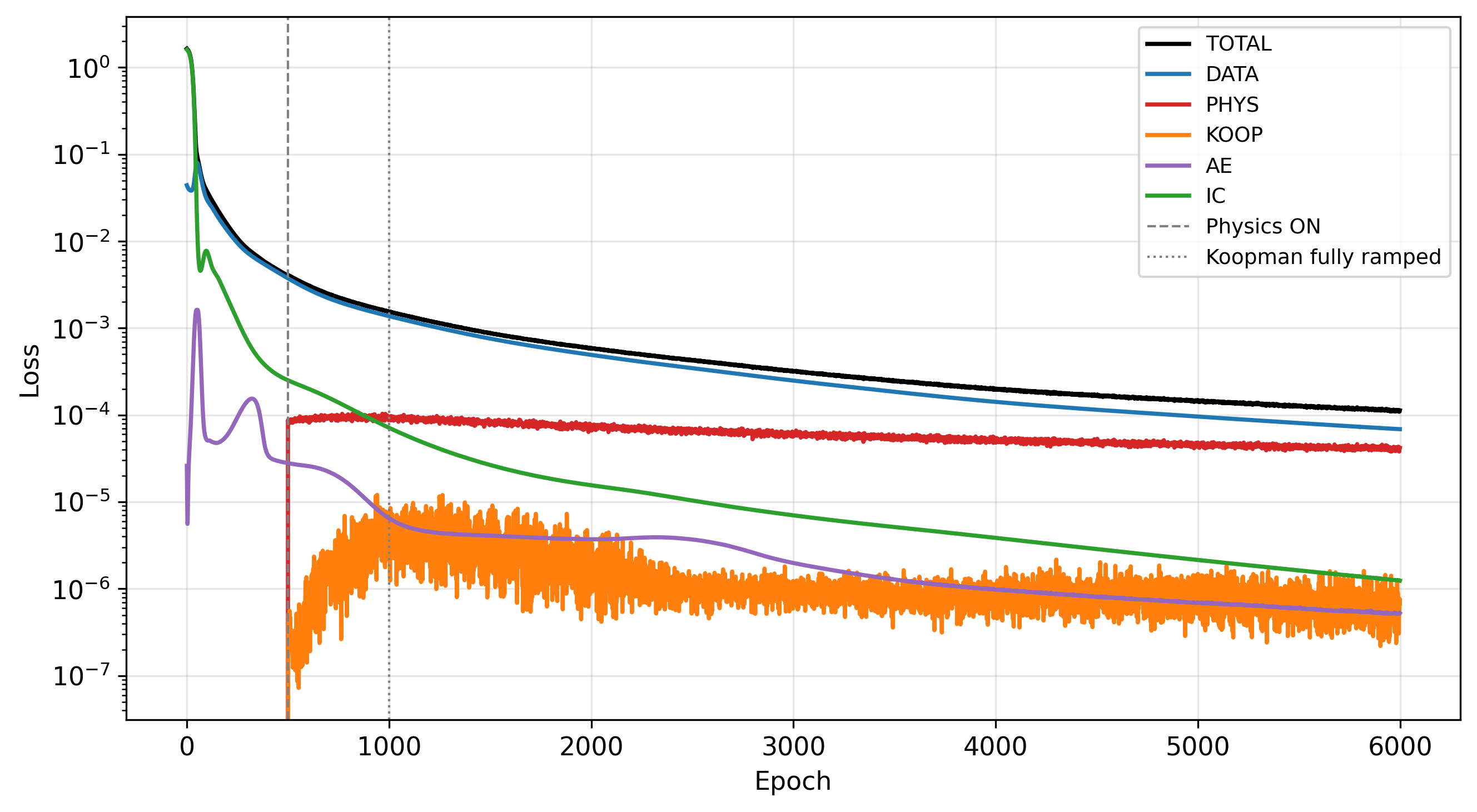}
\caption{Sweden}\label{FB4c}
\end{subfigure}
\caption{Evolution of the composite K--PINN loss function during training for the Germany, Morocco, and Sweden datasets.}\label{FB4}
\end{figure}

Figure~\ref{FB4} depicts the evolution of the composite K--PINN loss function during training. 
A consistent decrease of the loss is observed for all datasets, indicating stable optimization of the data reconstruction, Koopman consistency, and physics-informed residual terms. 

\begin{figure}[H]
\centering
\begin{subfigure}[b]{.5\textwidth}
\centering
\includegraphics[width=1\textwidth]{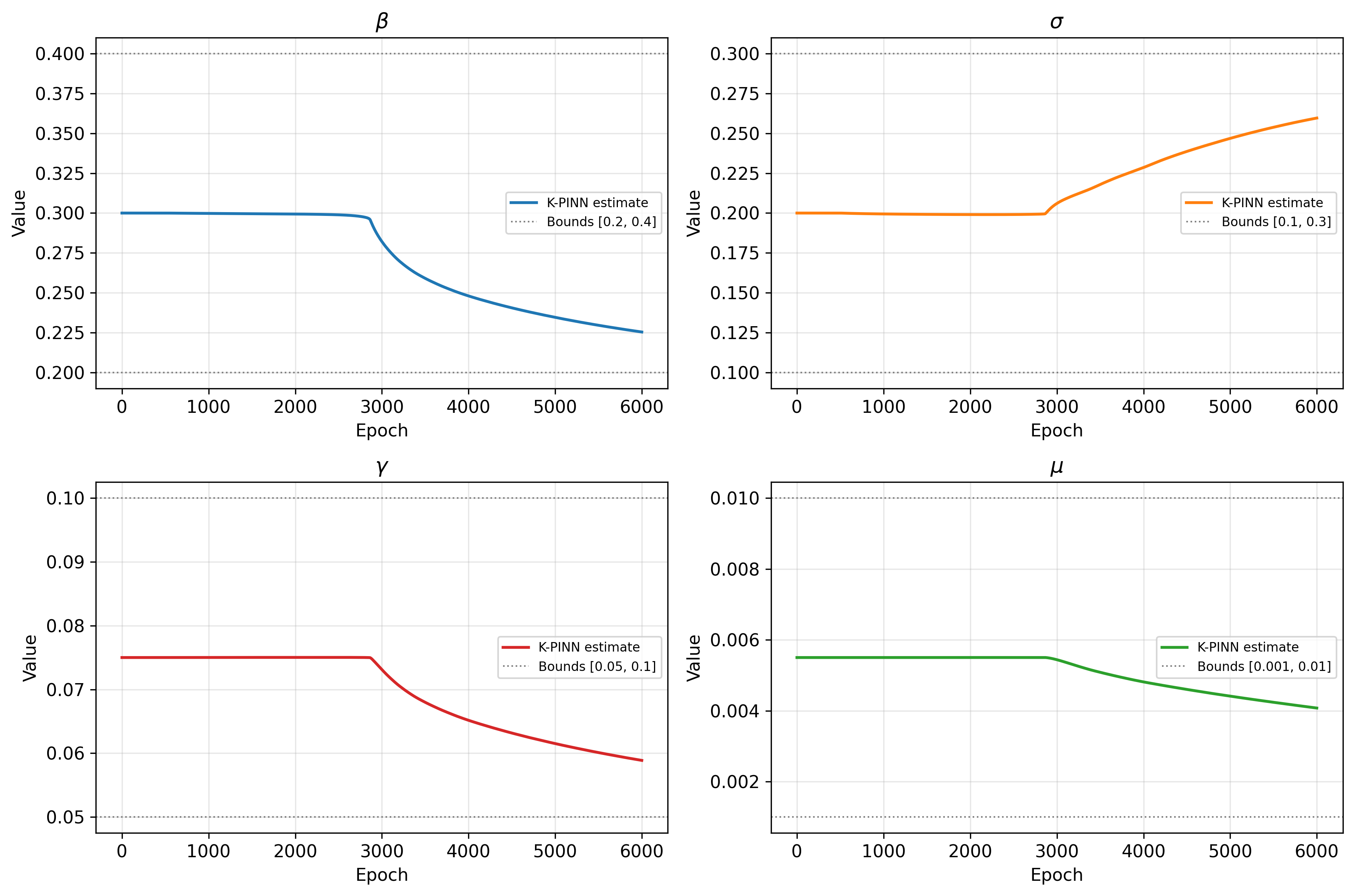}
\caption{Germany}\label{FB5a}
\end{subfigure}%
~
\begin{subfigure}[b]{.5\textwidth}
\centering
\includegraphics[width=1\textwidth]{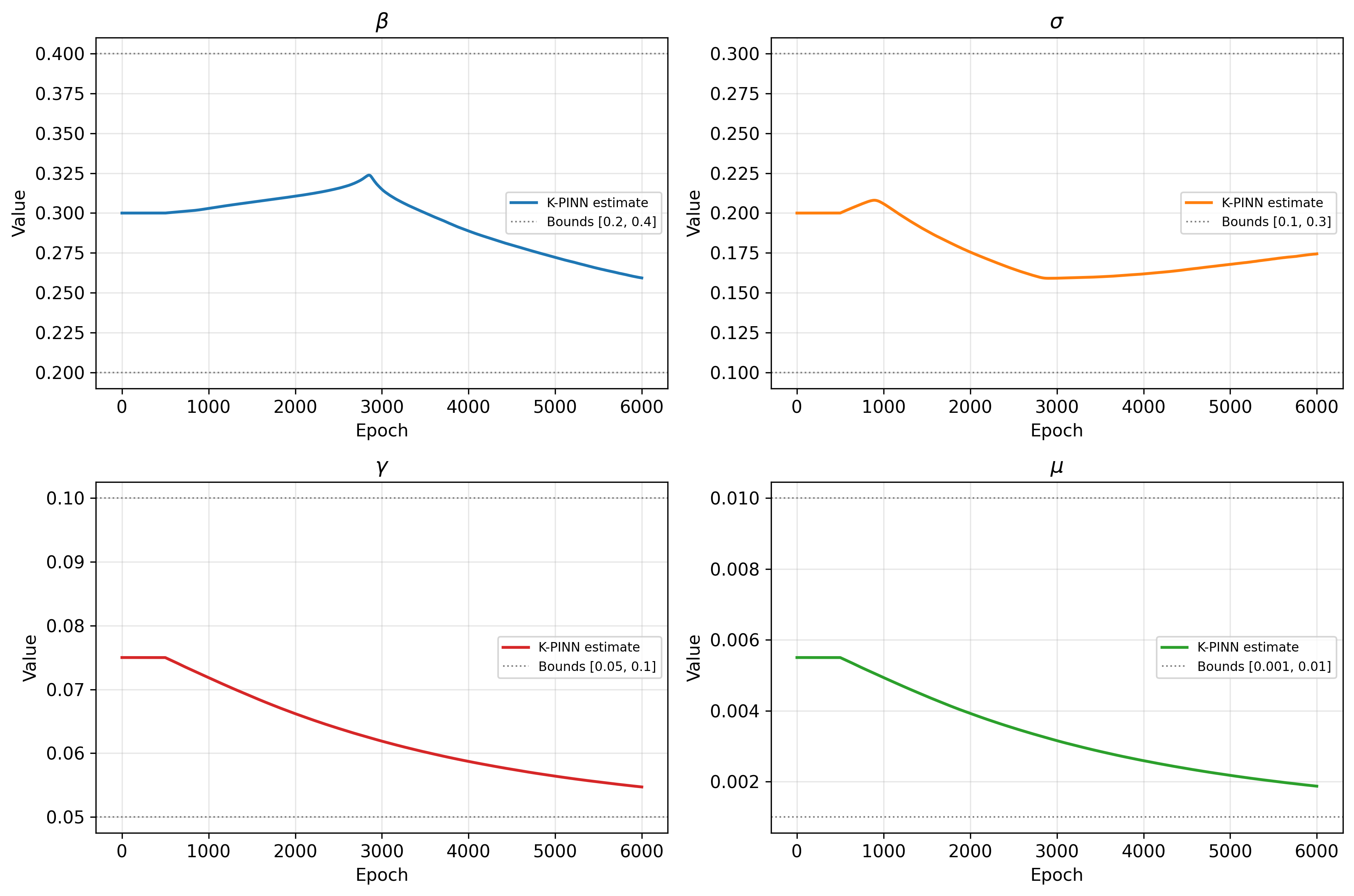}
\caption{Morocco}\label{FB5b}
\end{subfigure}
~
\begin{subfigure}[b]{.5\textwidth}
\centering
\includegraphics[width=\textwidth]{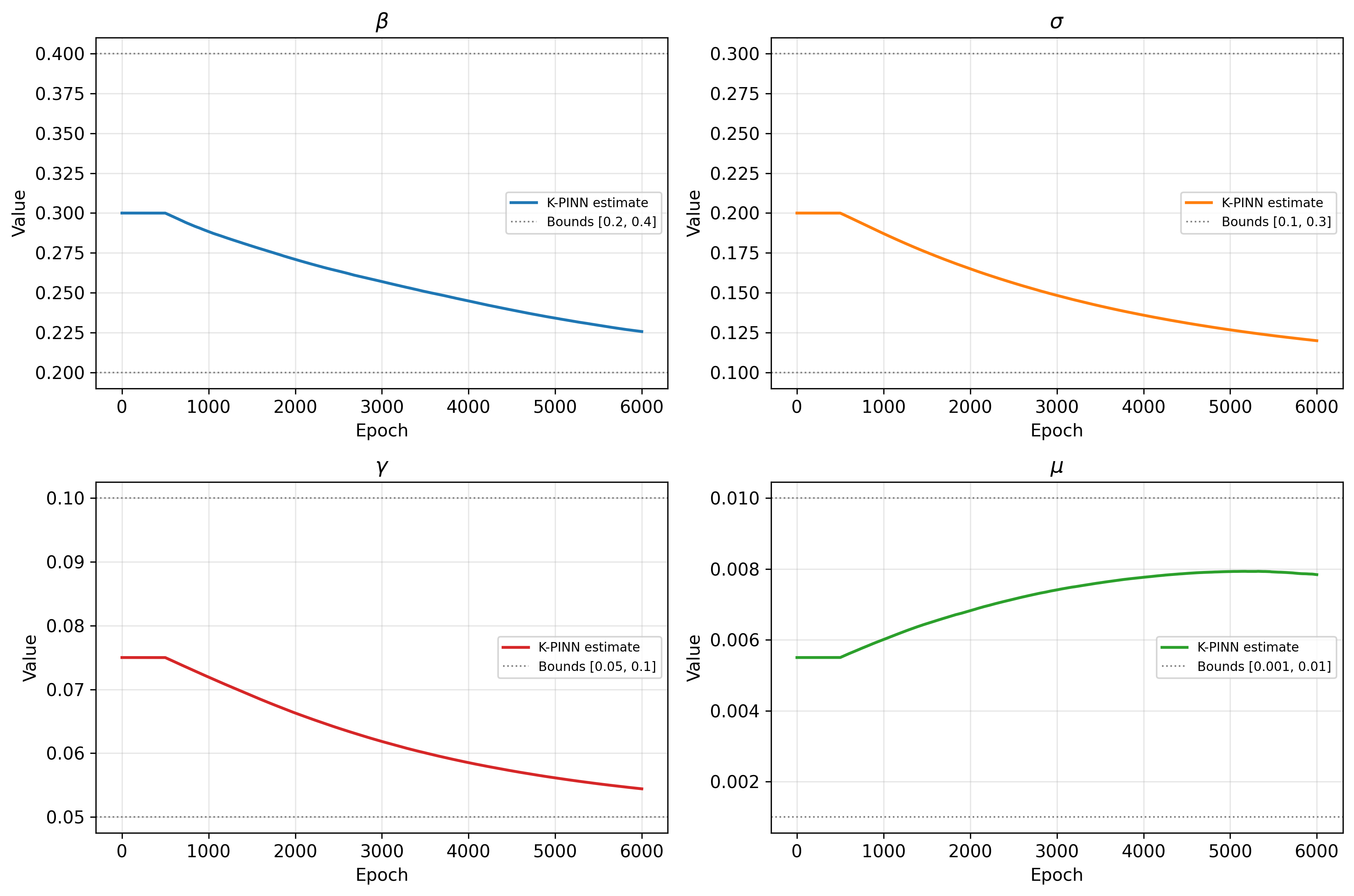}
\caption{Sweden}\label{FB54c}
\end{subfigure}
\caption{Convergence histories of the epidemiological parameter estimates obtained by the proposed K--PINN framework for the Germany, Morocco, and Sweden datasets.}\label{FB5}
\end{figure}

Figure~\ref{FB5} displays the evolution of the epidemiological parameter estimates learned by the proposed K--PINN framework. 
The estimated parameters smoothly converge toward stable values, which demonstrates the robustness of the inverse-learning strategy. 
The resulting parameter estimates are summarized in Table~\ref{Tab6} and are consistent with biologically plausible epidemic regimes, supporting the improved predictive performance of the proposed methodology.
\end{appendices}


\section*{Declarations}

\subsection*{Conflict of Interest} 
\noindent
The authors declared that they have no conflict of interest.


\subsection*{Author Contributions}
\noindent
A. Zinihi: Conceptualization, Soft\-ware, Methodology, Validation, Formal Analysis, Investigation, Writing-Original Draft, Writing-Review and Editing, Visualization.

\noindent
M. Ehrhardt: Conceptualization, Supervision, Methodology, Formal Analysis, Investigation, Writing-Review and Editing.

\noindent
M. R. Sidi Ammi: Conceptualization, Supervision, Methodology, Formal Analysis.



\subsection*{Computational Setup} 
\noindent
All numerical simulations for the K--EDMD, PINN, and K--PINN frameworks were conducted on a workstation running Microsoft Windows 11 Home. 
The system features an Intel Core Ultra 9 275HX CPU (24 cores, 24 threads) with 32 GB of RAM.\\
Computations were accelerated using a hybrid GPU setup, including an integrated Intel GPU and an NVIDIA GeForce RTX 5070 Ti Laptop GPU (12 GB VRAM).

\subsection*{Software Environment} 
\noindent
The simulations were implemented in \textit{Python 3.13} using PyTorch.

\subsection*{Data Availability} 
\noindent
This study uses both synthetic and real epidemic datasets. 
Real datasets were obtained
from the following publicly available source: \cite{WHO2024Mpox, Lancet2024, UKGov2025, WHO2024Covid, Gehrcke2025, OWD2025Estimated, worldometers}.

\bibliographystyle{unsrt}
\bibliography{References}

\end{document}